\pageno=0
\font\twelverm=cmr12
\font\elfrm=cmr10 scaled 1097
\font\tenrm=cmr10
\font\tentt=cmtt10
\font\tenit=cmti10
\font\tensl=cmsl10
\font\elfsl=cmsl10 scaled 1097
\font\ninerm=cmr9
\font\eightrm=cmr8
\font\sixrm=cmr6
\font\twelvei=cmmi12
\font\twelvebi=cmmib10 scaled\magstep1
\font\tenbi=cmmib10
\font\elfi=cmmi10 scaled 1097
\font\ninei=cmmi9
\font\eighti=cmmi8
\font\sixi=cmmi6
\font\twelvesy=cmsy10 scaled\magstep1
\font\elfsy=cmsy10 scaled 1097
\font\ninesy=cmsy9
\font\eightsy=cmsy8
\font\sixsy=cmsy6
\font\twelvebf=cmbx12
\font\elfbf=cmbx10 scaled 1097
\font\ninebf=cmbx9
\font\eightbf=cmbx8
\font\sixbf=cmbx6
\font\twelvett=cmtt12
\font\ninett=cmtt9
\font\eighttt=cmtt8
\font\twelveit=cmti12
\font\elfit=cmti10 scaled 1097
\font\nineit=cmti9
\font\eightit=cmti8
\font\twelvesl=cmsl12
\font\ninesl=cmsl9
\font\eightsl=cmsl8
\font\twelveex=cmex10 scaled\magstep1
\font\bsi=cmbsy10 scaled\magstep1
\font\si=cmbsy10
\newskip\ttglue
\def\twelvepoint{\def\rm{\fam0\twelverm}%
\def\sl{\fam\slfam\twelvesl}%
\textfont0=\twelverm   \scriptfont0=\ninerm  \scriptscriptfont0=\sevenrm%
\textfont1=\twelvei    \scriptfont1=\ninei   \scriptscriptfont1=\seveni%
\textfont2=\twelvesy   \scriptfont2=\ninesy  \scriptscriptfont2=\sevensy%
\textfont3=\twelveex   \scriptfont3=\twelveex\scriptscriptfont3=\twelveex%
\textfont\itfam=\twelveit    \def\it{\fam\itfam\twelveit}%
\textfont\slfam=\twelvesl    \def\sl{\fam\slfam\twelvesl}%
\textfont\ttfam=\twelvett    \def\tt{\fam\ttfam\twelvett}%
\textfont\bffam=\twelvebf   \scriptfont\bffam=\ninebf%
\scriptscriptfont\bffam=\sevenbf     \def\bf{\fam\bffam\twelvebf}%
\ttglue=.5em plus .25em minus.15em%
\normalbaselineskip=15pt\rm%
\setbox\strutbox=\hbox{\vrule height9.75pt depth4pt width0pt}}
\def\elfpoint{\def\rm{\fam0\elfrm}%
\def\sl{\fam\slfam\elfsl}%
\textfont0=\elfrm\scriptfont0=\ninerm\scriptscriptfont0=\sixrm%
\textfont1=\elfi\scriptfont1=\ninei\scriptscriptfont1=\sixi%
\textfont2=\elfsy\scriptfont2=\ninesy\scriptscriptfont2=\sixsy%
\textfont3=\tenex\scriptfont3=\tenex\scriptscriptfont3=\tenex%
\textfont\itfam=\elfit\def\it{\fam\itfam\elfit}%
\scriptscriptfont\bffam=\sixbf\def\bf{\fam\bffam\elfbf}%
\ttglue=.45em plus.25em minus.15em%
\normalbaselineskip=13.5pt\rm%
\setbox\strutbox=\hbox{\vrule height7.8pt depth3.15pt width0pt}}

\def\tenpoint{\def\rm{\fam0\tenrm}%
\def\sl{\fam\slfam\tensl}%
\def\tt{\fam\ttfam\tentt}%
\textfont0=\tenrm\scriptfont0=\sevenrm\scriptscriptfont0=\fiverm%
\textfont1=\teni\scriptfont1=\seveni\scriptscriptfont1=\fivei%
\textfont2=\tensy\scriptfont2=\sevensy\scriptscriptfont2=\fivesy%
\textfont3=\tenex\scriptfont3=\tenex\scriptscriptfont3=\tenex%
\textfont\itfam=\tenit\def\it{\fam\itfam\tenit}%
\textfont\ttfam=\tentt\def\tt{\fam\ttfam\tentt}%
\scriptscriptfont\bffam=\fivebf\def\bf{\fam\bffam\tenbf}%
\ttglue=.5em plus.25em minus.15em%
\normalbaselineskip=12.5pt\rm%
\setbox\strutbox=\hbox{\vrule height8.5pt depth3.5pt width0pt}}

\def\ninepoint{\def\rm{\fam0\ninerm}
 \textfont0=\ninerm   \scriptfont0=\sixrm \scriptscriptfont0=\fiverm
 \textfont1=\ninei    \scriptfont1=\sixi  \scriptscriptfont1=\fivei
 \textfont2=\ninesy   \scriptfont2=\sixsy \scriptscriptfont2=\fivesy
 \textfont3=\tenex    \scriptfont3=\tenex \scriptscriptfont3=\tenex
 \textfont\itfam=\nineit    \def\it{\fam\itfam\nineit}
 \textfont\slfam=\ninesl    \def\sl{\fam\slfam\ninesl}
 \textfont\ttfam=\ninett    \def\tt{\fam\ttfam\ninett}
 \textfont\bffam=\ninebf   \scriptfont\bffam=\sixbf
  \scriptscriptfont\bffam=\fivebf     \def\bf{\fam\bffam\ninebf}
\ttglue=.5em plus .25em minus.15em
\normalbaselineskip=11pt\rm
\setbox\strutbox=\hbox{\vrule height7.7pt depth3pt width0pt}}

\def\eightpoint{\def\rm{\fam0\eightrm}
 \textfont0=\eightrm  \scriptfont0=\sixrm  \scriptscriptfont0=\fiverm
 \textfont1=\eighti   \scriptfont1=\sixi   \scriptscriptfont1=\fivei
 \textfont2=\eightsy  \scriptfont2=\sixsy  \scriptscriptfont2=\fivesy
 \textfont3=\tenex    \scriptfont3=\tenex  \scriptscriptfont3=\tenex
 \textfont\itfam=\eightit  \def\it{\fam\itfam\eightit}
 \textfont\slfam=\eightsl  \def\sl{\fam\slfam\eightsl}
 \textfont\ttfam=\eighttt  \def\tt{\fam\ttfam\eighttt}
 \textfont\bffam=\eightbf  \scriptfont\bffam=\sixbf
  \scriptscriptfont\bffam=\fivebf  \def\bf{\fam\bffam\eightbf}
\ttglue=.5em plus.25em minus.15em
\normalbaselineskip=9pt\rm
\setbox\strutbox=\hbox{\vrule height7pt depth2pt width0pt}}

\font\rmb=cmr12 scaled \magstep 1
 1
\font\rmc=cmr12 scaled \magstep 2
\font\rmd=cmr12 scaled \magstep 3
\font\rme=cmr12 scaled \magstep 4
 2
\font\ita=cmti12
\font\itb=cmti12 scaled \magstep 1
 3
 4
 5
\font\mitb=cmmi10 scaled \magstep 2
\font\mitc=cmmi10 scaled \magstep 3
\font\mitd=cmmi10 scaled \magstep 4
 5

 1
\def\title#1#2#3{\par\ifnum\prevgraf>0 \npageno=0 \fi
   \global\shortno=0\global\secno=0\global\tabno=0\global
   \figno=0\global\footno=0\global\refno=0\goodbreak\ifnum\npageno=0
   \par\vskip 2cm plus2.5mm minus2.5mm\fi{\noindent\textfont1=\mitd
   \rmd #1$\vphantom{\hbox{\rmd y}}$}\par
   \penalty1000\vskip8pt\noindent{\itb #2$\vphantom{\hbox{\itb y}}$}{\par
   \penalty1000\vskip7.5pt\baselineskip11.5pt\hrule height 0.1pt\par
   \penalty1000\vskip7pt{\tenpoint\noindent\rm #3$\vphantom{y}$}\par
   \penalty1000\vskip6pt\hrule height 0.1pt
   \par\penalty1000\vskip9pt\noindent}}
\def\stitle#1#2#3{\par\ifnum\prevgraf>0 \npageno=0 \fi
   \global\shortno=0\global\secno=0\global\tabno=0\global
   \figno=0\global\footno=0\global\refno=0\goodbreak\ifnum\npageno=0
   \par\vskip 1.6cm plus2mm minus2mm\fi{\noindent\textfont1=\mitc
   \rmc #1$\vphantom{\hbox{\rmc y}}$}\par
   \penalty1000\vskip6.5pt\noindent{\ita #2$\vphantom{\hbox{\itb y}}$}{\par
   \penalty1000\vskip6pt\baselineskip10pt\hrule height 0.1pt\par
   \penalty1000\vskip5.75pt{\ninepoint\noindent\rm #3$\vphantom{y}$}\par
   \penalty1000\vskip4.8pt\hrule height 0.1pt
   \par\penalty1000\vskip7.25pt\noindent}}
\def\supertitle#1#2#3#4{\par\ifnum\prevgraf>0 \npageno=0 \fi
   \global\shortno=0\global\secno=0\global\tabno=0\global
   \figno=0\global\footno=0\global\refno=0\goodbreak\ifnum\npageno=0
   \par\vskip 1.3cm plus2.5mm minus2.5mm\fi\noindent{\textfont1=\mitc
   \rmc #1}\par
   \penalty1000\vskip1cm plus 3mm minus 7mm{\noindent\textfont1=\mitd\rmd
   #2$\vphantom{\hbox{\rmd y}}$}\par\penalty1000\vskip8pt\noindent{\itb
   #3$\vphantom{\hbox{\itb y}}$}{\par\penalty1000\vskip7.5pt\baselineskip
   11.5pt\hrule height 0.1pt\par\penalty1000\vskip7pt{\tenpoint\noindent\rm
   #4$\vphantom{y}$}\par\penalty1000\vskip6pt\hrule height 0.1pt\par
   \penalty1000\vskip9pt\noindent}}
\def\ssupertitle#1#2#3#4{\par\ifnum\prevgraf>0 \npageno=0 \fi
   \global\shortno=0\global\secno=0\global\tabno=0\global
   \figno=0\global\footno=0\global\refno=0\goodbreak\ifnum\npageno=0
   \par\vskip 1.1cm plus2.1mm minus2.1mm\fi\noindent{\textfont1=\mitb
   \rmb #1$\vphantom{\hbox{\rmb y}}$}\par
   \penalty1000\vskip8mm plus 3mm minus 6mm{\noindent\textfont1=\mitc\rmc
   #2$\vphantom{\hbox{\rmc y}}$}\par\penalty1000\vskip8pt\noindent{\ita
   #3$\vphantom{\hbox{\ita y}}$}{\par\penalty1000\vskip6pt\baselineskip
   10pt\hrule height
   0.1pt\par\penalty1000\vskip5.5pt{\ninepoint\noindent\rm
   #4$\vphantom{y}$}\par\penalty1000\vskip6pt\hrule height 0.1pt\par
   \penalty1000\vskip9pt\noindent}}
%
%
\def\firstreftitle#1#2#3#4{\par\ifnum\prevgraf>0 \npageno=0 \fi
   \global\shortno=0\global\secno=0\global\tabno=0\global
   \figno=0\global\footno=0\global\refno=0\goodbreak\ifnum\npageno=0
   \par\vskip 1.5cm plus2.5mm minus2.5mm\fi\noindent{\textfont1=\mitd
   \rmc Progress in Meteor Science}{\par\penalty1000
\vskip2mm plus 0.5mm minus 0.5mm\baselineskip11.5pt{\tenpoint\noindent\it
Articles in this section have been formally refereed by at least one
professional and one experienced, knowledgeable amateur meteor worker,
and deal with global analyses of meteor data, methods for meteor
observing and data reduction, observations with professional equipment, or
theoretical studies.}\par}\par\penalty1000
\vskip7.5mm plus 2.5mm minus 2.5mm{\noindent\textfont1=\mitd\rmd
#1$\vphantom{\hbox{\rmd y}}$}\par\penalty1000\vskip8pt\noindent{\itb
#2$\vphantom{\hbox{\itb y}}$}{\tenpoint\rm
\baselineskip11.5pt\footnote{}{\hskip-\parindent #3\smallnewpar\it
WGN, the Journal of the
International Meteor Organization, Vol.~\the\volno, No.~\the\nrno,
\month~\the\yearno, pp.~\the\firstpageno--\the\lastpageno.}}%
{\par\penalty1000\vskip7.5pt\baselineskip
11.5pt\hrule height 0.1pt\par\penalty1000\vskip7pt{\tenpoint\noindent\rm
#4$\vphantom{y}$}\par\penalty1000\vskip6pt\hrule height 0.1pt\par
\penalty1000\vskip9pt\noindent}}%
\def\reftitle#1#2#3#4{\par\ifnum\prevgraf>0 \npageno=0 \fi
   \global\shortno=0\global\secno=0\global\tabno=0\global
   \figno=0\global\footno=0\global\refno=0\goodbreak\ifnum\npageno=0
   \par\vskip 1.5cm plus2.5mm minus2.5mm\fi\noindent{\textfont1=\mitd\rmd
#1$\vphantom{\hbox{\rmd y}}$}\par\penalty1000\vskip8pt\noindent{\itb
#2$\vphantom{\hbox{\itb y}}$}{\tenpoint\rm
\baselineskip11.5pt\footnote{}{\hskip-\parindent #3\smallnewpar\it
WGN, the Journal of the
International Meteor Organization, Vol.~\the\volno, No.~\the\nrno/3,
\month~\the\yearno, pp.~\the\firstpageno--\the\lastpageno.}}%
{\par\penalty1000\vskip7.5pt\baselineskip
11.5pt\hrule height 0.1pt\par\penalty1000\vskip7pt{\tenpoint\noindent\rm
#4$\vphantom{y}$}\par\penalty1000\vskip6pt\hrule height 0.1pt\par
\penalty1000\vskip9pt\noindent}}%
\def\noabsupertitle#1#2#3{\par\ifnum\prevgraf>0 \npageno=0 \fi
   \global\shortno=0\global\secno=0\global\tabno=0\global
   \figno=0\global\footno=0\global\refno=0\goodbreak\ifnum\npageno=0
   \par\vskip 2cm plus2.5mm minus2.5mm\fi\noindent{\rmc #1}\par
   \penalty1000\vskip1cm plus 5mm minus 5mm{\noindent\rmd
   #2$\vphantom{\hbox{\rmd y}}$}\par\penalty1000\vskip8pt\noindent{\itb
   #3$\vphantom{\hbox{\itb y}}$}{\par\penalty1000\vskip7.5pt
   \hrule height 0.1pt\par\penalty1000\vskip9pt\noindent}}
\def\noabtitle#1#2{\par\ifnum\prevgraf>0 \npageno=0\fi
   \global\shortno=0\global\secno=0\global\tabno=0\global
   \figno=0\global\footno=0\global\refno=0\goodbreak\ifnum\npageno=0
   \par\vskip2cm plus 2.5mm minus 2.5mm\fi
   {\noindent\rmd #1$\vphantom{\hbox{\rme y}}$}\par\penalty1000\vskip8pt
   \noindent{\itb #2$\vphantom{\hbox{\itb y}}$}\par\penalty1000\vskip7.5pt
   \hrule height0.1pt\par\penalty1000\vskip9pt\noindent}
\def\snoabtitle#1#2{\par\ifnum\prevgraf>0 \npageno=0\fi
   \global\shortno=0\global\secno=0\global\tabno=0\global
   \figno=0\global\footno=0\global\refno=0\goodbreak\ifnum\npageno=0
   \par\vskip1.6cm plus 2mm minus 2mm\fi
   {\noindent\textfont1=\mitc
   \rmc #1$\vphantom{\hbox{\rmc y}}$}\par\penalty1000\vskip6.5pt
   \noindent{\ita #2$\vphantom{\hbox{\ita y}}$}\par\penalty1000\vskip6pt
   \hrule height0.1pt\par\penalty1000\vskip7.25pt\noindent}
\def\shorttitle#1#2{\global\secno=0\global\tabno=0\global
   \figno=0\global\footno=0\global\refno=0{\noindent
   \rmd #1$\vphantom{\hbox{\rme y}}$}\par
   \penalty1000\vskip8pt\noindent{\itb #2$\vphantom{\hbox{\itb y}}$}\par
   \penalty1000\vskip7.5pt\hrule height0.1pt\par\penalty1000
   \vskip9pt\noindent}
\def\sshorttitle#1#2{\global\secno=0\global\tabno=0\global
   \figno=0\global\footno=0\global\refno=0{\noindent
   \rmc #1$\vphantom{\hbox{\rmd y}}$}\par
   \penalty1000\vskip6.5pt\noindent{\ita #2$\vphantom{\hbox{\ita y}}$}\par
   \penalty1000\vskip6pt\hrule height0.1pt\par\penalty1000
   \vskip7.25pt\noindent}
\def\noauttitle#1{\global\secno=0\global\tabno=0\global\figno=0
 \global\footno=0\global\refno=0{\noindent\rmd #1$\vphantom{\hbox{\rmd y}}$}
   \par\vskip 7.5pt\hrule height
   0.1pt\par\penalty1000\vskip9pt\noindent}
\def\noabsemibigtitle#1#2#3#4{\global\shortno=0\global\secno=0
   \global\tabno=0\global\figno=0\global\footno=0\global\refno=0\goodbreak
   \ifnum\npageno=0 \par\vskip 2cm plus2.5mm  minus2.5mm\fi{\noindent
   \rmc #1$\vphantom{\hbox{\rmd y}}$}\par
   \penalty1000\vskip7pt\noindent{\rmd #2$\vphantom{\hbox{\rmc y}}$}\par
   \penalty1000\vskip7pt\noindent{\rmd #3$\vphantom{\hbox{\rmc y}}$}\par
 \par\penalty1000\vskip7pt\noindent{\itb #4$\vphantom{\hbox{\itb y}}$}{\par
   \penalty1000\vskip7.5pt\baselineskip11.5pt\hrule height 0.1pt\par
   \penalty1000\vskip9pt\noindent}}
\def\supersemititle#1#2#3#4#5{\global\secno=0\global\tabno=0\global\figno=0
   \global\footno=0\global\refno=0\goodbreak\ifnum\npageno=0
   \par\vskip 1.3cm plus2.5mm minus2.5mm\fi\noindent{\textfont1=\mitd
   \rmc #1$\vphantom{\hbox{\rmc y}}$}\par
   \penalty1000\vskip7pt\noindent{\textfont1=\mitd\rmd
   #2$\vphantom{\hbox{\rmd y}}$}\par
   \penalty1000\vskip7pt\noindent{\rmd #3$\vphantom{\hbox{\rmd y}}$}\par
 \par\penalty1000\vskip8pt\noindent{\itb #4$\vphantom{\hbox{\itb y}}$}{\par
   \penalty1000\vskip7.5pt\baselineskip11.5pt\hrule height 0.1pt\par
   \penalty1000\vskip7pt{\tenpoint\noindent\rm
   #5$\vphantom{y}$}\par\penalty1000\vskip6pt\hrule height 0.1pt\par
   \penalty1000\vskip9pt\noindent}}
\def\noabsupersemititle#1#2#3#4{\global\secno=0\global\tabno=0\global\figno=0
   \global\footno=0\global\refno=0\goodbreak\ifnum\npageno=0
   \par\vskip 1.3cm plus2.5mm minus2.5mm\fi\noindent{\textfont1=\mitd
   \rmc #1$\vphantom{\hbox{\rmc y}}$}\par
   \penalty1000\vskip7pt\noindent{\textfont1=\mitd\rmd
   #2$\vphantom{\hbox{\rmd y}}$}\par
   \penalty1000\vskip7pt\noindent{\rmd #3$\vphantom{\hbox{\rmd y}}$}\par
 \par\penalty1000\vskip8pt\noindent{\itb #4$\vphantom{\hbox{\itb y}}$}{\par
   \penalty1000\vskip7.5pt\baselineskip11.5pt\hrule height 0.1pt\par
   \penalty1000\vskip9pt\noindent}}
\def\ssupersemititle#1#2#3#4#5{\global\secno=0\global\tabno=0\global\figno=0
   \global\footno=0\global\refno=0\goodbreak\ifnum\npageno=0
   \par\vskip 1.3cm plus2.5mm minus2.5mm\fi\noindent{\textfont1=\mitb
   \rmb #1$\vphantom{\hbox{\rmb y}}$}\par
   \penalty1000\vskip6.5pt\noindent{\rmc #2$\vphantom{\hbox{\rmc y}}$}\par
   \penalty1000\vskip6.5pt\noindent{\rmb #3$\vphantom{\hbox{\rmb y}}$}\par
 \par\penalty1000\vskip8pt\noindent{\ita #4$\vphantom{\hbox{\ita y}}$}{\par
   \penalty1000\vskip7pt\baselineskip10.35pt\hrule height 0.1pt\par
   \penalty1000\vskip6pt{\ninepoint\noindent\rm
   #5$\vphantom{y}$}\par\penalty1000\vskip5.75pt\hrule height 0.1pt\par
   \penalty1000\vskip8pt\noindent}}

\def\bigtitle#1#2#3#4{\par\ifnum\prevgraf>0 \npageno=0 \fi
   \global\shortno=0\global\secno=0\global\tabno=0\global
   \figno=0\global\footno=0\global\refno=0\goodbreak\ifnum\npageno=0
   \par\vskip 2cm plus2.5mm  minus2.5mm\fi{\noindent\textfont1=\mitd\rmd #1$
   \vphantom{\hbox{\rmd y}}$}\par\penalty1000\vskip8pt\noindent
   {\textfont1=\mitd\rmd #2$\vphantom{\hbox{\rmd y}}$}
\par\penalty1000\vskip 8pt\noindent{\itb #3$\vphantom{\hbox{\itb y}}$}{\par
   \penalty1000\vskip7.5pt\baselineskip11.5pt\hrule height 0.1pt\par
   \penalty1000\vskip 7pt{\tenpoint\noindent\rm #4$\vphantom{y}$}\par
   \penalty1000\vskip6pt\hrule height 0.1pt\par\vskip9pt\noindent}}
\def\noabbigtitle#1#2#3{\global\shortno=0\global\secno=0\global\tabno=0\global
   \figno=0\global\footno=0\global\refno=0\goodbreak\ifnum\npageno=0
   \par\vskip 2cm plus2.5mm  minus2.5mm\fi{\noindent
   \rmd #1$\vphantom{\hbox{\rmd y}}$}\par
   \penalty1000\vskip8pt\noindent{\rmd #2$\vphantom{\hbox{\rmd y}}$}
 \par\penalty1000\vskip 8pt\noindent{\itb #3$\vphantom{\hbox{\itb y}}$}{\par
   \penalty1000\vskip7.5pt\hrule height 0.1pt\par
   \penalty1000\vskip9pt\noindent}}
\def\snoabbigtitle#1#2#3{\global\shortno=0
   \global\secno=0\global\tabno=0\global
   \figno=0\global\footno=0\global\refno=0\goodbreak\ifnum\npageno=0
   \par\vskip 1.6cm plus2mm  minus2mm\fi{\noindent\textfont1=\mitc%
   \rmc #1$\vphantom{\hbox{\rmc y}}$}\par
   \penalty1000\vskip6pt\noindent{\rmc #2$\vphantom{\hbox{\rmc y}}$}
 \par\penalty1000\vskip 6pt\noindent{\ita #3$\vphantom{\hbox{\ita y}}$}{\par
   \penalty1000\vskip6pt\hrule height 0.1pt\par
   \penalty1000\vskip7.25pt\noindent}}
\def\snoabsemititle#1#2#3{\global\shortno=0
   \global\secno=0\global\tabno=0\global
   \figno=0\global\footno=0\global\refno=0\goodbreak\ifnum\npageno=0
   \par\vskip 1.6cm plus2mm  minus2mm\fi{\noindent\textfont1=\mitc%
   \rmc #1$\vphantom{\hbox{\rmc y}}$}\par
   \penalty1000\vskip6pt\noindent{\rmb #2$\vphantom{\hbox{\rmb y}}$}
   \par\penalty1000\vskip 6pt\noindent{\ita #3$\vphantom{\hbox{\ita y}}$}{\par
   \penalty1000\vskip6pt\hrule height 0.1pt\par
   \penalty1000\vskip7.25pt\noindent}}
\def\snoabsupertitle#1#2#3{\global\shortno=0
   \global\secno=0\global\tabno=0\global
   \figno=0\global\footno=0\global\refno=0\goodbreak\ifnum\npageno=0
   \par\vskip 1.6cm plus2mm  minus2mm\fi{\noindent\textfont1=\mitb%
   \rmb #1$\vphantom{\hbox{\rmb y}}$}\par
   \penalty1000\vskip6pt\noindent{\rmc #2$\vphantom{\hbox{\rmc y}}$}
   \par\penalty1000\vskip 6pt\noindent{\ita #3$\vphantom{\hbox{\ita y}}$}{\par
   \penalty1000\vskip6pt\hrule height 0.1pt\par
   \penalty1000\vskip7.25pt\noindent}}
\def\snoabbigsemititle#1#2#3#4{\global\shortno=0
   \global\secno=0\global\tabno=0\global
   \figno=0\global\footno=0\global\refno=0\goodbreak\ifnum\npageno=0
   \par\vskip 1.6cm plus2mm  minus2mm\fi{\noindent
   \rmc #1$\vphantom{\hbox{\rmc y}}$}\par
   \penalty1000\vskip6pt{\noindent
   \rmc #2$\vphantom{\hbox{\rmc y}}$}\par
   \penalty1000\vskip6pt\noindent{\rmb #3$\vphantom{\hbox{\rmb y}}$}
 \par\penalty1000\vskip 6pt\noindent{\ita #4$\vphantom{\hbox{\ita y}}$}{\par
   \penalty1000\vskip6pt\hrule height 0.1pt\par
   \penalty1000\vskip7.25pt\noindent}}
\def\semititle#1#2#3#4{\global\shortno=0\global\secno=0\global\tabno=0\global
   \figno=0\global\footno=0\global\refno=0\goodbreak\ifnum\npageno=0
   \par\vskip 2cm plus2.5mm  minus2.5mm\fi{\noindent
   \rmd #1$\vphantom{\hbox{\rmd y}}$}\par
   \penalty1000\vskip7pt\noindent{\rmc #2$\vphantom{\hbox{\rmc y}}$}
 \par\penalty1000\vskip 7pt\noindent{\itb #3$\vphantom{\hbox{\itb y}}$}{\par
   \penalty1000\vskip7.5pt\baselineskip11.5pt\hrule height 0.1pt\par
   \penalty1000\vskip 7pt{\tenpoint\noindent\rm #4$\vphantom{y}$}\par
   \penalty1000\vskip6pt\hrule height 0.1pt\par\vskip9pt\noindent}}
\def\noabsemititle#1#2#3{\global\shortno=0\global\secno=0\global\tabno=0
    \global\figno=0\global\footno=0\global\refno=0\goodbreak
    \ifnum\npageno=0 \par\vskip 2cm plus2.5mm  minus2.5mm\fi{\noindent
    \rmd #1$\vphantom{\hbox{\rmd y}}$}\par
   \penalty1000\vskip7pt\noindent{\rmc #2$\vphantom{\hbox{\rmc y}}$}
 \par\penalty1000\vskip7pt\noindent{\itb #3$\vphantom{\hbox{\itb y}}$}{\par
   \penalty1000\vskip7.5pt\baselineskip11.5pt\hrule height 0.1pt\par
   \penalty1000\vskip9pt\noindent}}
\def\noautshortnote#1{\global\secno=0\global\tabno=0\global
   \figno=0\global\footno=0\global\refno=0\goodbreak\ifnum\npageno=0
   \ifnum\shortno=0 \vskip6mm plus 2mm minus 2mm\noindent
   \hrule height 1.2pt\par\vskip8.5mm plus 2.5mm minus 2.5mm
   \else\vskip17mm plus 4mm minus 3mm\fi\fi
   \global\shortno=1\noindent\noauttitle{#1}}
\def\shortnote#1#2{\global\secno=0\global\tabno=0\global
   \figno=0\global\footno=0\global\refno=0\goodbreak\ifnum\npageno=0
   \ifnum\shortno=0 \vskip6mm plus 2mm minus 2mm\noindent
   \hrule height 1.2pt\par\vskip8.5mm plus 2.5mm minus 2.5mm
   \else\vskip17mm plus 4mm minus 3mm\fi\fi
   \global\shortno=1\noindent\shorttitle{#1}{#2}}
\def\sshortnote#1#2{\global\secno=0\global\tabno=0\global
   \figno=0\global\footno=0\global\refno=0\goodbreak\ifnum\npageno=0
   \ifnum\shortno=0 \vskip5mm plus 1.75mm minus 1.75mm\noindent
   \hrule height 1.2pt\par\vskip7.25mm plus 2mm minus 2mm
   \else\vskip15.5mm plus 3.5mm minus 2.5mm\fi\fi
   \global\shortno=1\noindent\sshorttitle{#1}{#2}}
\def\shortnotes#1#2{\global\secno=0\global\tabno=0\global
   \figno=0\global\footno=0\global\refno=0\goodbreak\ifnum\npageno=0
   \ifnum\shortno=0 \vskip8.5mm plus 2.5mm minus 2.5mm
   \noindent\hrule height 1.2pt\par\vskip8.5mm plus 2.5mm minus 2.5mm
   \else\vskip17mm plus 5mm minus 5mm\fi\fi
 \global\shortno=1\noindent{\rmc Short Notes}\par\penalty10000\vskip8.5mm
    plus 2.5mm minus 2.5mm\noindent\shorttitle{#1}{#2}}
\parskip=0pt
\parindent=25pt
\def\newpar{\edef\next{\hangafter=\the\hangafter
    \hangindent=\the\hangindent}
    \par\penalty10000\ifnum\prevgraf>0 \global\npageno=0\fi
    \par\penalty-100\vskip 4pt plus 2pt minus 2pt\next%
    \edef\next{\ifnum\prevgraf>-\the\hangafter\prevgraf=0\hangafter=1
         \hangindent=0pt\else\prevgraf=\the\prevgraf\fi}
    \noindent\next}
\def\smallnewpar{\edef\next{\hangafter=\the\hangafter
    \hangindent=\the\hangindent}
    \par\penalty10000\ifnum\prevgraf>0 \global\npageno=0\fi
    \par\penalty-100\vskip 3pt plus 1.5pt minus 1pt\next%
    \edef\next{\ifnum\prevgraf>-\the\hangafter\prevgraf=0\hangafter=1
         \hangindent=0pt\else\prevgraf=\the\prevgraf\fi}
    \noindent\next}
\def\pensmallnewpar{\edef\next{\hangafter=\the\hangafter
    \hangindent=\the\hangindent}
    \par\penalty10000\ifnum\prevgraf>0 \global\npageno=0\fi
    \par\penalty10000\vskip 3pt plus 1.5pt minus 1pt\next%
    \edef\next{\ifnum\prevgraf>-\the\hangafter\prevgraf=0\hangafter=1
         \hangindent=0pt\else\prevgraf=\the\prevgraf\fi}
    \noindent\next}
\newcount\npageno
\newcount\shortno
\newcount\secno
\newcount\tabno
\newcount\figno
\newcount\refno
\newdimen\figindent
\newdimen\figheight
\newdimen\newfigheight
\newdimen\figwidth
\npageno=0
\shortno=0
\secno=0
\refno=0
\tabno=0
\figno=0
\figindent=0sp
\figheight=0sp
\newfigheight=0sp
\figwidth=0sp
\def\section#1{\par\global\npageno=0
    \global\advance\secno by 1\ifnum\prevgraf=0 \relax
    \else\penalty-100\vskip 12pt plus 6pt minus 4.5pt\fi
    {\textfont0=\twelvebf\textfont1=\twelvebi\textfont2=\bsi
     \scriptfont0=\ninebf
    \noindent{\bf\the\secno.\ #1}}\par\penalty10000
    \vskip 4pt plus 2pt minus2pt\noindent}%
\def\smallsection#1{\par\global\npageno=0
    \global\advance\secno by 1\ifnum\prevgraf=0 \relax
    \else\penalty-100\vskip 10pt plus 5pt minus 4pt\fi
    {\textfont0=\tenbf\textfont1=\tenbi\textfont2=\si
    \noindent{\bf\the\secno.\ #1}}\newpar}
\def\appsection#1{\par\global\npageno=0
   \ifnum\prevgraf=0 \relax
   \else\penalty-100\vskip 12pt plus 6pt minus 4.5pt\fi
   {\textfont0=\twelvebf\textfont1=\twelvebi\textfont2=\bsi
   \noindent{\bf#1}}\newpar}
\def\smallappsection#1{\par\global\npageno=0
   \ifnum\prevgraf=0 \relax
   \else\penalty-100\vskip 10pt plus 5pt minus 4pt\fi
   {\textfont0=\tenbf\textfont1=\tenbi\textfont2=\si
   \noindent{\bf#1}}\newpar}
\def\artref#1#2#3#4{\par\global\advance\refno by 1\ifnum\prevgraf=0 \relax
    \else\penalty-100\vskip 2pt plus 0.5pt minus 0.5pt\fi
    \hangindent\parindent\noindent\hbox to \parindent{[$
    \the\refno$]\hfill}#1, ``#2'', {\it #3\/}, #4.}
\def\sartref#1#2#3#4{\par\global\advance\refno by 1\ifnum\prevgraf=0 \relax
    \else\penalty-100\vskip 1.5pt plus 0.5pt minus 0.5pt\fi
    \hangindent\parindent\noindent\hbox to \parindent{[$
    \the\refno$]\hfill}#1, ``#2'', {\it #3\/}, #4.}
\def\noautref#1#2#3{\par\global\advance\refno by 1\ifnum\prevgraf=0 \relax
    \else\penalty-100\vskip 2pt plus 0.5pt minus 0.5pt\fi
    \hangindent\parindent\noindent\hbox to \parindent{[$
    \the\refno$]\hfill}``#1'', {\it #2\/}, #3.}
\def\noartref#1#2#3{\par\global\advance\refno by 1\ifnum\prevgraf=0 \relax
    \else\penalty-100\vskip 2pt plus 0.5pt minus 0.5pt\fi
    \hangindent\parindent\noindent\hbox to \parindent{[$
    \the\refno$]\hfill}#1, {\it #2\/}, #3.}
\def\snoartref#1#2#3{\par\global\advance\refno by 1\ifnum\prevgraf=0 \relax
    \else\penalty-100\vskip 1.5pt plus 0.5pt minus 0.5pt\fi
    \hangindent\parindent\noindent\hbox to \parindent{[$
    \the\refno$]\hfill}#1, {\it #2\/}, #3.}
\def\noartnoautref#1#2{\par\global\advance\refno by 1\ifnum\prevgraf=0 \relax
    \else\penalty-100\vskip 2pt plus 0.5pt minus 0.5pt\fi
    \hangindent\parindent\noindent\hbox to \parindent{[$
    \the\refno$]\hfill}{\it #1\/}, #2.}
\def\bookref#1#2#3{\par\global\advance\refno by 1\ifnum\prevgraf=0 \relax
    \else\penalty-100\vskip 2pt plus 0.5pt minus 0.5pt\fi
    \hangindent\parindent\noindent\hbox to \parindent{[$
    \the\refno$]\hfill}#1, ``#2'', #3.}
\def\sbookref#1#2#3{\par\global\advance\refno by 1\ifnum\prevgraf=0 \relax
    \else\penalty-100\vskip 1.5pt plus 0.5pt minus 0.5pt\fi
    \hangindent\parindent\noindent\hbox to \parindent{[$
    \the\refno$]\hfill}#1, ``#2'', #3.}
\def\noautbookref#1#2{\par\global\advance\refno by 1\ifnum\prevgraf=0 \relax
    \else\penalty-100\vskip 2pt plus 0.5pt minus 0.5pt\fi
    \hangindent\parindent\noindent\hbox to \parindent{[$
    \the\refno$]\hfill}``#1'', #2.}
\def\shortref#1#2{\par\global\advance\refno by 1\ifnum\prevgraf=0 \relax
    \else\penalty-100\vskip 2pt plus 0.5pt minus 0.5pt\fi
    \hangindent\parindent\noindent\hbox to \parindent{[$
    \the\refno$]\hfill}#1, ``#2''.}
\def\sourceref#1{\par\global\advance\refno by 1\ifnum\prevgraf=0 \relax
    \else\penalty-100\vskip 2pt plus 0.5pt minus 0.5pt\fi
    \hangindent\parindent\noindent\hbox to \parindent{[$
    \the\refno$]\hfill}{\it #1.}}
\def\perscom#1#2{\par\global\advance\refno by 1\ifnum\prevgraf=0 \relax
    \else\penalty-100\vskip 2pt plus 0.5pt minus 0.5pt\fi
    \hangindent\parindent\noindent\hbox to \parindent{[$
    \the\refno$]\hfill}#1, {\it personal communications\/}, #2.}
\def\libref#1#2{\par\hangindent\parindent\noindent
   \hbox to \parindent{$\bullet$\hfill}{\it #1}\pensmallnewpar{\rm #2}\newpar}

\newbox\tabbox
\def\table#1#2{\global\npageno=0\advance\tabno by 1
    \setbox\tabbox=\vbox{\tenpoint\tabskip=0pt
    \offinterlineskip\halign{#2}}
       $$\vbox{\tenpoint\baselineskip11.5pt\ialign{\hfill$##$\crcr
                  \vtop{\hsize=\wd\tabbox\hangindent1.65truecm\noindent%
                  \hbox to 1.65truecm{Table \the\tabno\ -- \hfill}#1}\crcr
                  \noalign{\vskip8pt}
                  \box\tabbox\crcr}}$$}
\def\nocaptable#1{\global\npageno=0\advance\tabno by 1
    \setbox\tabbox=\vbox{\tenpoint\tabskip=0pt
    \offinterlineskip\halign{#1}}
       $$\vbox{\tenpoint\baselineskip11.5pt\ialign{\hfill$##$\crcr
                  \box\tabbox\crcr}}$$
}\def\smalltable#1#2{\global\npageno=0\advance\tabno by 1
    \setbox\tabbox=\vbox{\ninepoint\tabskip=0pt
    \offinterlineskip\halign{#2}}
       $$\vbox{\ninepoint\baselineskip10.35pt\ialign{\hfill$##$\crcr
                  \vtop{\hsize=\wd\tabbox\hangindent1.5truecm\noindent%
      \hbox to 1.5truecm{\ninerm Table \the\tabno\ -- \hfill}#1}\crcr
                  \noalign{\vskip8pt}
                  \box\tabbox\crcr}}$$}
\def\nocapsmalltable#1{\global\npageno=0\advance\tabno by 1
    \setbox\tabbox=\vbox{\ninepoint\tabskip=0pt
    \offinterlineskip\halign{#1}}
       $$\vbox{\ninepoint\baselineskip10.35pt\ialign{\hfill$##$\crcr
                  \box\tabbox\crcr}}$$}
\def\bigtable#1#2{\global\npageno=0\advance\tabno by 1
    \setbox\tabbox=\vbox{\tabskip=0pt
    \offinterlineskip\halign{#2}}
       $$\vbox{\ialign{\hfill$##$\crcr
                  \vtop{\hsize=\wd\tabbox\hangindent2truecm\noindent%
                  \hbox to 2truecm{Table \the\tabno\ -- \hfill}#1}\crcr
                  \noalign{\vskip9.5pt}
                  \box\tabbox\crcr}}$$}
\newbox\figbox
\newcount\hgtno
\newcount\basno
\def\figure#1#2#3#4#5{\par\global\advance\figno by 1 \ifnum\prevgraf=0
    \noindent
    \else\newpar\fi \figindent=#1\advance\figindent by 5mm%
    \setbox\figbox=\vbox{\tenpoint\baselineskip11.5pt\hsize
        =#1\hangindent1.75truecm\noindent\hbox to 1.75truecm{Figure \the\figno
        \ -- \hfill}#5}%
    \figheight=#2\advance\figheight by 16pt\advance\figheight by \ht\figbox
    \hgtno=\figheight \basno=\baselineskip \divide\hgtno by \basno
    \advance\hgtno by 1 \advance\hgtno by #3 \advance\figheight by #4%
    \hangindent=\the\figindent\hangafter=-\the\hgtno\noindent
    \hskip-\the\figindent$\smash{\hbox to \the\figindent{\vtop to
    \the\figheight{\hsize=#1\vfill\vskip8pt\vbox{\tenpoint
    \baselineskip11.5pt\hsize=#1\hangindent1.75truecm\noindent
    \hbox to 1.75truecm{Figure \the\figno\ --
    \hfill}#5}\vskip8pt\ }\hfill}}$}
\def\smallboxfigure#1#2#3#4#5{\par\global\advance\figno by 1
\ifnum\prevgraf=0
    \noindent
    \else\smallnewpar\fi \figindent=#1\advance\figindent by 5mm%
    \figwidth=#1\advance\figwidth by -2.4pt
    \setbox\figbox=\vbox{\ninepoint\baselineskip10.35pt\rm\hsize
     =#1\hangindent1.65truecm\noindent\hbox to 1.65truecm{Figure \the\figno
        \ -- \hfill}#5}%
    \figheight=#2\advance\figheight by 16pt\advance\figheight by \ht\figbox
    \hgtno=\figheight \basno=\baselineskip \divide\hgtno by \basno
    \advance\hgtno by 1 \advance\hgtno by #3 \advance\figheight by #4%
    \hangindent=\the\figindent\hangafter=-\the\hgtno
    \newfigheight=#2\advance\newfigheight by -2.4pt
    \noindent \hskip-\the\figindent$\smash{\hbox to \the\figindent{\vtop to
    \the\figheight{\vfill\hsize=#1\vskip-3pt\ialign to
    #1{\vrule##width1.2pt&\hfill\hbox to
    \figwidth{\hfil##\hfil}\hfill&\vrule##width1.2pt\crcr
    \noalign{\hrule height1.2pt}%
    height\newfigheight&\ &height\newfigheight\crcr
    \noalign{\hrule height1.2pt}}\vfill\vskip8pt\vbox{
   \ninepoint\baselineskip10.35pt\rm\hsize=#1\hangindent1.65truecm\noindent
    \hbox to 1.65truecm{Figure \the\figno\ --\hfill}#5}\vskip0pt\
    }\hfill}}$}
\def\nocapsmallfigure#1#2#3#4{\par\global\advance\figno by 1 \ifnum\prevgraf=0
    \noindent
    \else\smallnewpar\fi \figindent=#1\advance\figindent by 5mm%
    \figheight=#2\advance\figheight by 8pt
    \hgtno=\figheight \basno=\baselineskip \divide\hgtno by \basno
    \advance\hgtno by 1 \advance\hgtno by #3 \advance\figheight by #4%
    \hangindent=\the\figindent\hangafter=-\the\hgtno\noindent
    \hskip-\the\figindent$\smash{\hbox to \the\figindent{\vtop to
    \the\figheight{\hsize=#1\vfill}\hfill}}$}
\def\fullfigure#1#2#3#4{\par\global\npageno=0
    \global\advance\figno by 1 \ifnum\prevgraf=0 \noindent
    \else\newpar\fi
    \setbox\figbox=\vbox{\tenpoint\baselineskip11.5pt\hsize
    =#1\hangindent1.75truecm\noindent\hbox to 1.75truecm{Figure \the\figno
        \ -- \hfill}#3}%
    \figheight=#2\advance\figheight by 16pt\advance\figheight by \ht\figbox
    $$\vbox to \figheight{\hsize=#1\vfill#4\vskip8pt\vbox{
     \tenpoint\baselineskip11.5pt\hsize=#1\hangindent1.75truecm\noindent
     \hbox to 1.75truecm{Figure \the\figno\ --\hfill}#3}\vskip2.5pt}$$}
\def\nocapfullfigure#1#2#3#4{\par\global\npageno=0
    \global\advance\figno by 1 \ifnum\prevgraf=0 \noindent
    \else\newpar\fi
    \setbox\figbox=\vbox{\tenpoint\baselineskip11.5pt\hsize
    =#1\noindent #3}%
    \figheight=#2\advance\figheight by 16pt\advance\figheight by \ht\figbox
    $$\vbox to \figheight{\hsize=#1\vfill#4\vskip8pt\vbox{
     \tenpoint\baselineskip11.5pt\hsize=#1\noindent #3}\vskip2.5pt}$$}
\def\plaktable#1#2#3{\par\global\npageno=0
    \global\advance\tabno by 1 \ifnum\prevgraf=0 \noindent
    \else\newpar\fi
    \setbox\figbox=\vbox{\tenpoint\baselineskip11.5pt\hsize
    =#1\hangindent1.75truecm\noindent\hbox to 1.75truecm{Table
    \the\tabno\ -- \hfill}#3}%
    \figheight=#2\advance\figheight by 16pt\advance\figheight by \ht\figbox
    $$\vbox to \figheight{\hsize=#1\vbox{\tenpoint\baselineskip11.5pt
     \hsize=#1\hangindent1.75truecm\noindent
     \hbox to 1.75truecm{Table \the\tabno\ --\hfill}#3}\vfill}$$}
\def\fullboxfigure#1#2#3{\par\global\npageno=0
    \global\advance\figno by 1 \ifnum\prevgraf=0 \noindent
    \else\newpar\fi
\figwidth=#1\advance\figwidth by -2.4pt
    \setbox\figbox=\vbox{\tenpoint\baselineskip11.5pt\hsize
    =#1\hangindent1.75truecm\noindent\hbox to 1.75truecm{Figure \the\figno
        \ -- \hfill}#3}%
    \figheight=#2\advance\figheight by 24pt\advance\figheight by
\ht\figbox
\newfigheight=#2\advance\newfigheight by -2.4pt
    $$\vbox to
\figheight{\vfill\vskip8pt\ialign to #1{\vrule##width1.2pt&\hfill\hbox to \figwidth{\hfil##\hfil}\hfill&\vrule##width1.2pt\crcr
\noalign{\hrule height1.2pt}
height\newfigheight&\ &height\newfigheight\crcr
\noalign{\hrule height1.2pt}}\vfill\vskip8pt\vbox{
     \tenpoint\baselineskip11.5pt\hsize=#1\hangindent1.75truecm\noindent
     \hbox to 1.75truecm{Figure \the\figno\
     --\hfill}#3}\vskip2.5pt}$$}
\def\smallfullboxfigure#1#2#3{\par\global\npageno=0
    \global\advance\figno by 1 \ifnum\prevgraf=0 \noindent
    \else\newpar\fi
\figwidth=#1\advance\figwidth by -2.4pt
    \setbox\figbox=\vbox{\ninepoint\baselineskip10.35pt\hsize
    =#1\hangindent1.65truecm\noindent\hbox to 1.65truecm{Figure \the\figno
        \ -- \hfill}#3}%
    \figheight=#2\advance\figheight by 24pt\advance\figheight by
\ht\figbox
\newfigheight=#2\advance\newfigheight by -2.4pt
    $$\vbox to
\figheight{\vfill\vskip8pt\ialign to
#1{\vrule##width1.2pt&\hfill\hbox to
\figwidth{\hfil##\hfil}\hfill&\vrule##width1.2pt\crcr
\noalign{\hrule height1.2pt}
height\newfigheight&\ &height\newfigheight\crcr
\noalign{\hrule height1.2pt}}\vfill\vskip8pt\vbox{
     \ninepoint\baselineskip10.35pt\hsize=#1\hangindent1.65truecm\noindent
     \hbox to 1.65truecm{Figure \the\figno\ --\hfill}#3}\vskip2.5pt}$$}
\def\smallfullfigure#1#2#3{\par\global\npageno=0
    \global\advance\figno by 1 \ifnum\prevgraf=0 \noindent
    \else\newpar\fi
    \setbox\figbox=\vbox{\ninepoint\baselineskip10.35pt\hsize
     =#1\hangindent1.65truecm\noindent\hbox to 1.65truecm{Figure \the\figno
        \ -- \hfill}#3}%
    \figheight=#2\advance\figheight by 16pt\advance\figheight by \ht\figbox
    $$\vbox to \figheight{\hsize=#1\vfill\vskip8pt\vbox{
     \ninepoint\baselineskip10.35pt\hsize=#1\hangindent1.65truecm\noindent
     \hbox to 1.65truecm{Figure \the\figno\ --\hfill}#3}\vskip2.5pt}$$}
\def\nocapsmallfullfigure#1#2#3{\par\global\npageno=0
    \global\advance\figno by 1 \ifnum\prevgraf=0 \noindent
    \else\newpar\fi
    \setbox\figbox=\vbox{\ninepoint\baselineskip10.35pt\hsize
     =#1{#3}}%
    \figheight=#2\advance\figheight by 16pt\advance\figheight by \ht\figbox
    $$\vbox to \figheight{\hsize=#1\vfill\vskip8pt\vbox{
     \ninepoint\baselineskip10.35pt\hsize=#1{\noindent #3}}\vskip2.5pt}$$}

\newskip\basskip
\basskip=11.5pt
\def\racco#1{\hbox to 10pt{$\smash{\vcenter{\hbox{$\left.\vphantom{\vcenter{
\vrule height #1\basskip}}\right\}$}}\hfill}$}}
\def\raccol#1{$\smash{\raise 0.5\basskip\hbox{#1}}$}

\def\0{\phantom{0}}
\def\1{\phantom{.0}}
\def\E{\hbox to 6truemm{\hfill E\hfill}}
\def\W{\hbox to 6truemm{\hfill W\hfill}}
\def\N{\hbox to 6truemm{\hfill N\hfill}}
\def\S{\hbox to 6truemm{\hfill S\hfill}}
\mathchardef\g="020E
\mathchardef\h="0068
\mathchardef\m="006D
\mathchardef\s="0073
\newbox\help
\newbox\punt
\newskip\terug
\newskip\vooruit
\def\dg{{}^\g\setbox\help=\hbox{${}^\g$}\setbox\punt=\hbox{$.$}\skip
\terug=-.5\wd\help plus0pt minus0pt\advance\skip\terug by -0.17em plus0em
minus0em\hskip\skip\terug\skip\vooruit=-\skip\terug\advance\skip\vooruit by
-\wd\punt.\hskip\skip\vooruit}
\def\dh{{}^\h\setbox\help=\hbox{${}^\h$}\setbox\punt=\hbox{$.$}\skip
\terug=-.5\wd\help plus0pt minus0pt\advance\skip\terug by -0.17em plus0em
minus0em\hskip\skip\terug\skip\vooruit=-\skip\terug\advance\skip\vooruit by
-\wd\punt.\hskip\skip\vooruit}
\def\dm{{}^\m\setbox\help=\hbox{${}^\m$}\setbox\punt=\hbox{$.$}\skip
\terug=-.5\wd\help plus0pt minus0pt\advance\skip\terug by -0.17em plus0em
minus0em\hskip\skip\terug\skip\vooruit=-\skip\terug\advance\skip\vooruit by
-\wd\punt.\hskip\skip\vooruit}
\def\ds{{}^\s\setbox\help=\hbox{${}^\s$}\setbox\punt=\hbox{$.$}\skip
\terug=-.5\wd\help plus0pt minus0pt\advance\skip\terug by -0.17em plus0em
minus0em\hskip\skip\terug\skip\vooruit=-\skip\terug\advance\skip\vooruit by
-\wd\punt.\hskip\skip\vooruit}
\def\dpr{{}^\prime\setbox\help=\hbox{${}^\prime$}\setbox\punt=\hbox{$.$}\skip
\terug=-.5\wd\help plus0pt minus0pt\advance\skip\terug by -0.17em plus0em
minus0em\hskip\skip\terug\skip\vooruit=-\skip\terug\advance\skip\vooruit by
-\wd\punt.\hskip\skip\vooruit}
\def\ddp{{}^{\prime\prime}\setbox\help=\hbox{${}^{\prime\prime}$}
\setbox\punt=\hbox{$.$}\skip
\terug=-.5\wd\help plus0pt minus0pt\advance\skip\terug by -0.17em plus0em
minus0em\hskip\skip\terug\skip\vooruit=-\skip\terug\advance\skip\vooruit by
-\wd\punt.\hskip\skip\vooruit}
\def\?{\hbox{\rm ?}}%
\def\advancepageno{\global\npageno=1\global\advance\pageno by 1\ifodd\pageno
    \global\hoffset=-1truecm\else\global\hoffset=0truecm\fi}
\ifodd\pageno\global\hoffset=0truecm\else\global\hoffset=-1truecm\fi
\def\makeheadline{\vbox to 0pt{\vskip-1.3truecm
     \line{\vbox to 9.75pt{}\the\headline}\vss}\nointerlineskip}
\newcount\volno
\newcount\nrno
\newcount\yearno
\headline={\elfpoint\sl\ifodd\pageno WGN, the Journal of the IMO
\the\volno:\the\nrno\ (\the\yearno)\hfill$\the\pageno$\else
$\the\pageno$\hfill WGN, the Journal of the IMO
\the\volno:\the\nrno\ (\the\yearno)\fi}
\newcount\footno
\global\footno=0
\def\nfoot#1{\global\npageno=0\advance\footno by 1
    {\tenpoint\baselineskip11.5pt\footnote{$^{\the\footno}$}{#1}}}
\def\smallnfoot#1{\global\npageno=0\advance\footno by 1%
{\ninepoint\baselineskip10.35pt\footnote{$^\the\footno$}{#1}}}%
\def\boxit#1{\vtop{\hrule height1.2pt \hbox{\vrule width1.2pt\kern3.5pt
    \vbox{\kern3.5pt#1\kern3.5pt}
    \kern3.5pt\vrule width1.2pt}\hrule height1.2pt}}
\hsize=17.2truecm
\hfuzz=2pt
\vfuzz=2pt
\vsize=25.2truecm
\voffset=-0.5truecm
\newcount\firstpageno
\newcount\lastpageno
\firstpageno=\pageno
\twelvepoint
\baselineskip 13.75pt
\nopagenumbers


\yearno=0000
\volno=00
\nrno=0
\title{Global Electrophonic Fireball Survey:\smallnewpar
       a review of witness reports - I.}%
{D. Vinkovi\'{c}, S. Garaj, P. L. Lim, D. Kova\v{c}i\'{c}, G. Zgrabli\'{c}, \v{Z}. Andrei\'{c} }%
{Despite more than 300 years since its first scientific description, the phenomenon of
electrophonic sounds from meteors are still eluding complete physical explanation.
According to the accepted knowledge, the sound itself is created by strong electric
fields on the ground induced by the meteor. Nonetheless, there is no convincing theory
that can fully explain how a meteor can generate such a strong electric field. Extreme
rareness of the phenomenon has prevented a substantial experimental work so far; thus,
consequently, it remains on the margins of scientific interest. This is quite
unfortunate since these electric fields suggest existence of a highly complex
electromagnetic coupling and charge dynamics between the meteors and the ionosphere.
Therefore, the existing theoretical work relies mostly on the witness reports. The
Global Electrophonic Fireball Survey (GEFS) is the first systematic survey of witness
reports of these sounds with a standardized questionnaire designed exclusively for this
phenomenon. Here we present the overall picture of the phenomenon that emerged after
almost 100 reports collected by GEFS. It becomes clear now that the lover meteor
brightness limit is about -2$^m$, suggesting a bias in the existing electrophonic sounds
catalogues toward brighter meteors. In contrast to the current belief that such low
brightness electrophonic meteors produce transient sounds, we find that they can also
produce sustained sounds. The current theories can not accommodate these results. We
revive the old idea that the electrophonic sounds can be created by the {\it corona
discharge} mechanism, in addition to the existing prevalent suggestion of resonant
vibration of objects on the ground.
}%

\section{Introduction}%
Audible sounds from meteors can be divided into two groups: {\it normal} and {\it
anomalous} ({\it electrophonic}) sounds. Normal sounds are acoustic waves produced
either by a hypersonic shock front or by a terminal burst and they propagate at the
speed of sound. Hence they display a noticeable time delay between the visual appearance
of the meteor and an audible detection on the ground. In contrast, anomalous sounds lack
this time delay, which means that the light and the sound are observed simultaneously.
The exact mechanism of their production by a meteor is still not known due to the
extreme rareness of this phenomenon.

\newpar
The first written record of distinction between the normal and anomalous sounds dates
back to the 17th century. Even though the concept of electromagnetic (EM) waves was
unknown at that time, almost instantaneous propagation of the anomalous sounds over a
large distance was suspected to be somehow connected to the ``electric matter''.
Nevertheless, the existence and reality of anomalous sounds was often denied by
scientists, especially when the real nature of meteors was discovered in the 19th
century. Since then, these sounds have been mainly ignored by the scientific community,
despite the persistent emergence of witness accounts. Consequently, the anomalous meteor
sounds have become the oldest unexplained astronomical phenomenon.

\newpar
Over the years, scarce theoretical research has managed to establish a connection
between the EM waves and anomalous sounds. In the first extensive study of these sounds,
Romig \& Lamar (1963) concluded that these sounds are most probably similar to the {\it
brontophonic sounds} (simultaneous with, or slightly preceding, the lightning stroke)
and {\it aurora sounds} (another poorly studied phenomenon --- sounds simultaneous with
bright auroras). They concluded that the sound is created by {\it corona discharge} on
sharp conductors, including plant leaves. Keay (1980) narrowed the frequency region for
these EM waves to the ELF/VLF (between 30Hz and 3kHz) region. He also conducted
experiments on human subjects and concluded that the ELF/VLF electric fields are capable
of entangling ordinary objects around the observer, from metals to dielectrics, into a
resonant vibration which then produces a sound in the same frequency range as the EM
waves (Keay \& Ostwald 1991). This has become a widely accepted theory and the corona
discharge mechanism has been mainly forgotten.

\par\eject\noindent
The term ``electrophonic sound'' was used for the first time in 1937 as a description
for sensation of a sound caused by electrical current through the head (Stevens 1937). A
few years later the term ``electrophonic bolide'' entered meteor astronomy as a
description of a bright meteor accompanied by anomalous sound (Dravert 1940).

\newpar
All this, however, merely moved the problem from how to create a sound to how to create
a strong ELF/VLF radiation from a meteor. Sound can be created from the ELF/VLF waves
only if the electric field at the ground is of at least several hundred V/m. Considering
the large distance between the observers on the ground and a meteor, the electric fields
in the vicinity of the meteor should be many magnitudes larger than on the ground (due
to distance square dependence for frequencies of kHz and higher and approximately
exponential dependence on distance for lower frequencies) (Wang, Tuan \& Silverman
1984). This problem has been studied theoretically by several authors (for an older
review see Bronshten 1991) (Beech \& Foschini 1999), including Keay (1980). Nonetheless,
the first instrumental recording of the electrophonic sounds combined with a video and
VLF observations during the 1998 Leonids showed that none of the existing theories can
explain the data (Zgrabli\'{c} et al 2002).

\newpar
Even though all these theories have been based solely on witness reports, there has been
no attempt to collect them with a standardized questionnaire. The existing catalogues of
electrophonic sound reports (Romig \& Lamar 1963 , Kaznev 1994, Keay 1993a) are usually
extracted from other sources, mainly from the fireball catalogues, and then
statistically analyzed. Considering how little we understand the nature behind this
phenomenon, the witness reports are still a valuable source of information. Two years
ago, we initiated the Global Electrophonic Fireball Survey (GEFS) to collect these
reports in a standardized form (Vinkovi\'c{} et al 2000). After receiving almost 100
reports, we present here a review of the collected data. Some reports of special
interest (such as the Leonids or meteors with complete trajectory) are presented in more
details. Due to the limited space in the journal, we can not present the complete
reports, but they can be accessed at the GEFS web-page {\it http://www.gefsproject.org}
or obtained from us by a request.

\section{Statistical analysis of the reported electrophonic sounds}

Witness reports were collected by the following methods: through the on-line HTML data
submission form, e-mail using a text version of the form, or informal e-mails. All of
them were transformed into the standardized survey form described by Vinkovi\'{c} et al
(2000). A single report often includes more than one person. Before we started preparing
this review, we received 91 reports of electrophonic meteors.

\newpar
The reports are designated as {\bf GEFSYYYY\_MM\_DD\_NN}, where YYYY\_MM\_DD is the date
of the electrophonic event (year, month, day) and NN is numeration in a case of more
than one event in a day. We consider one location as one event, no matter how many
observers are involved. If the auditory perception of electrophonic sounds does not
differ among observers, they will have more or less the same psychophysical reaction
regarding the sound description when exposed simultaneously to the same sound. Therefore
the sound is considered as one event instead of being interpreted as several events
based on the perception of multiple observers.

\newpar
 The geographical locations of the electrophonic meteors include: Australia, Belgium,
Canada, Croatia, Denmark, England, Finland, France, Germany, Israel, Mexico, Mongolia,
The Netherlands, Norway, Scotland, Singapore, Sweden, and the USA. The oldest event is
from the year 1952. The complete trajectory is calculated in three occasions. It is
interesting to note that 34 reports are associated with the Leonids and 7 with the
Perseids. Among them, there is one very interesting account of numerous electrophonic
sounds from the 1966 Leonids. In addition to the reports of electrophonic sounds from
meteors, there are three reports of (most probably) aurora sounds (GEFS1964\_11\_00\_02,
GEFS2001\_11\_23\_01 and GEFS2001\_12\_14\_01) and one report of an electrophonic sound
from the Space Shuttle reentry (mission STS-109) over central Texas (San Antonio). These
four reports are not included in the analysis shown below.

\par\eject\noindent
Here we statistically evaluate the data for specific segments of the GEFS form. We would
like to emphasize that most of the reports have very valuable information provided in
the {\it additional remarks} section of the form.

\newpar
{\bf 2.1. Personal information}

\newpar
The GEFS reports are sometimes not submitted by the witnesses themselves, but rather by
a person who collected various reports of sighted meteors and recognized reports of
electrophonic sounds among them. Such reports are sources of the events with known
meteor trajectory described in the next section; thus, the name of person who submitted
the data is not necessarily the name listed under {\it personal information}. If the
specific permission to use the witnesses' name as a reference to the submitted GEFS data
was not obtained then their name is omitted. If one GEFS report contains several
qualitatively different witness reports, the word ``multiple'' is used as personal
information and the names (or initials) are provided in conjunction with their GEFS
data. The level of meteor observing experience among the witnesses varies from {\it not
experienced} to {\it highly experienced}. Most witnesses had never heard a sound from a
meteor before, as expected.

\newpar
{\bf 2.2. Description of the observing site}

\newpar
The location of observing sites is usually described as a geographical feature, thus the
given coordinates correspond to these features and are not precise. The meteorological
conditions are described as {\it clear sky} and {\it calm} (windless or light breeze) in
84\% of reports with provided weather conditions. This is not surprising since such
conditions increase the possibility of spotting a meteor and noticing an unusual sound.

\newpar
{\bf 2.3. Details about the sound from the meteor}

\newpar
The exact month of the electrophonic event is provided in 73 reports but the day in only
47 (mainly because of very old events). The time is specified in 70 reports, usually an
estimate of the hour, thus only 32 reports have specified minutes or better.

\newpar
Descriptions of reported electrophonic sounds are matching descriptions in the existing
electrophonic catalogs (e.g. Keay 1993a, Kaznev 1994). Keay (1993b) classified the
electrophonic sounds into three groups: {\it smooth} (with 71\% rate of occurrence),
{\it staccato} (18\%), and {\it sharp} (11\%). This classification applied to the GEFS
reports is shown in Table~1. Our rate of occurrence of smooth and staccato sounds is
different, more than expected from Keay (1993b). This is probably due to different
methods used for counting sound events.

\newpar
In addition, some observers may not hear the sound or agree on its duration or
direction. The reported duration of sounds varies from less than a second to more than
10 seconds. Sound is recognized as coming from {\it all directions} in 19 reports
(27\%), {\it no direction} in 10 (14\%), and from {\it the meteor} in 41 (59\%). Air is
often mentioned as the direction or source of the sound. Three reports
(GEFS1998\_11\_16\_02, GEFS1998\_11\_17\_04, GEFS2001\_11\_18\_08) have an exact object
identified as a possible sound source.

\newpar
In 76 (84\%) cases, the meteors were spotted simultaneously with their sound. Observers
can not decide about a specific meteor that produced the sound in 8 (9\%) cases because
of high meteor activity. The electrophonic meteors were spotted prior to the sound in 2
(2\%) cases, but the sounds did not exceed the duration of their meteor. In 5 (5\%)
reports, the meteors were spotted after the sound. In two of such cases
(GEFS1972\_00\_00\_01, GEFS1969\_06\_00\_01), the electrophonic sound prompted the
observers to look toward the sky.

\newpar
Correlation of the sound with the meteor's light maximum reveals that: in 29 reports
witnesses {\it can not decide}, 48 (76\%) reports indicate {\it simultaneous} sound and
light maximum, 6 (10\%) reports indicate a sound {\it before}, and 9 (14\%) {\it after}
the light maximum (one report has two sounds with different correlations). Since some
reports deal with multiple sounds with the same type of correlation, the percentages
shown here suffer from large error bars.

\par\eject\noindent

 \table{{\it Phenomenological classification of electrophonic sounds. The
percentage shows the rate of occurrence in the GEFS catalog. According to Keay (1993b),
the sounds can be classified as} smooth, staccato, {\it and} sharp. {\it This
classification would correlate the sound frequency and duration with the meteor ELF/VLF
radiation of the same frequency and duration. In our study, we consider the possibility
of corona discharge as a source of some electrophonic sounds and apply different
classification according to two mechanisms of sound production:} vibration {\it or}
discharge. }
 {
  \vrule# &
  \quad\strut#\hfill\quad &
  \vrule# &
  \quad\strut#\hfill\quad &
  \vrule# &
  \quad\strut#\hfill\quad &
  \vrule#\cr
 \noalign{\hrule}
 height4pt&\omit&&\omit&&\omit &\cr
          &sound type&& rate && sound description &\cr
 height4pt&\omit&&\omit&&\omit&\cr
 \noalign{\hrule}
 height2pt width0pt& \omit &width0pt&\omit&width0pt&\omit&width0pt\cr
 \noalign{\hrule}
 height4pt& \omit &width0pt&\omit&width0pt&\omit&\cr
          & \omit&width0pt&\omit&width0pt&{\it classification according to Keay (1993b)}&\cr
 height4pt& \omit &width0pt&\omit&width0pt&\omit&\cr
 \noalign{\hrule}
 height4pt&\omit&&\omit&&\omit&\cr
          &smooth && 40.5\% && hissing, buzzing, whuss, whoosh, fizzing, bottle rocket, sjhh,&\cr
          &\omit && \omit && pchiu, steam escaping from cooker, sss, swishing, voom, &\cr
          &\omit && \omit &&  high-pitched whistle, whispering, sheewu &\cr
 height4pt&\omit&&\omit&&\omit&\cr
 \noalign{\hrule}
 height4pt&\omit&&\omit&&\omit&\cr
          & staccato && 47.0\%&& rustling, crackling, wood burning, phtt - like
                               electric arc, sizzling,&\cr
          &\omit && \omit &&  white noise,  shaking bulb with broken filament, zzz,  firework,&\cr
          &\omit && \omit &&  frying bacon, tzz, foam being ripped, like static, lit match, thrumming, &\cr
          &\omit && \omit &&  small single engine 'Cesna' airplane,   butter in hot pan,&\cr
          &\omit && \omit &&  hot metal in water, cards being shuffled, ice breaking up, electric flutter&\cr
 height4pt&\omit&&\omit&&\omit&\cr
 \noalign{\hrule}
 height4pt&\omit&&\omit&&\omit&\cr
          & sharp && 12.5\% && pop, thwuck, tic, boom, whump, clap, kweik &\cr
 height4pt&\omit&&\omit&&\omit&\cr
 \noalign{\hrule}
 height2pt width0pt& \omit &width0pt&\omit&width0pt&\omit&width0pt\cr
 \noalign{\hrule}
 height4pt& \omit &width0pt&\omit&width0pt&\omit&\cr
          & \omit&width0pt&\omit&width0pt&{\it classification according to our study}&\cr
 height4pt& \omit &width0pt&\omit&width0pt&\omit&\cr
 \noalign{\hrule}
 height4pt&\omit&&\omit&&\omit&\cr
          &vibration && 51.3\% && hissing, buzzing, fizzing, whuss, pop, thwuck, sjhh, tzz, bottle rocket, &\cr
          &\omit && \omit && shaking bulb with broken filament, sss, tic, steam escaping from cooker&\cr
          &\omit && \omit && high-pitched whistle, swishing, small single engine 'Cesna' airplane,&\cr
          &\omit && \omit && whispering, thrumming, boom, whump, voom, sheewu, clap, kweik&\cr
 height4pt&\omit&&\omit&&\omit&\cr
 \noalign{\hrule}
 height4pt&\omit&&\omit&&\omit&\cr
          & discharge && 48.7\%&& rustling, sizzling, whoosh, crackling, white noise,
                               `htt - like electric arc&\cr
          &\omit && \omit &&  wood burning, firework, frying bacon, zzz, pchiu, foam being ripped,&\cr
          &\omit && \omit &&  lit match, butter in hot pan, like static, hot metal in water,  &\cr
          &\omit && \omit &&  cards being shuffled, ice breaking up, electric flutter&\cr
 height4pt&\omit&&\omit&&\omit&\cr
 \noalign{\hrule}
 }

\newpar
In a case of one Perseid meteor, fading of the meteor's trail is described to correlate
with the loudness of a sizzling sound that ended with a ``pop'' (GEFS1995\_08\_10\_01).

\newpar
Two out of 6 reports of sound before the light maximum are actually marked as 'can not
decide', but their audio/video recordings show the sound preceding the final meteor
flash (GEFS1998\_11\_
17\_04 and GEFS1998\_11\_17\_05). This demonstrates that it is very
hard for an observer to make such a time estimate. These two recordings belong to the
1998 Leonids and show that a meteor can induce an electrophonic sound when it has
altitude of $\sim$100km (Zgrabli\'{c} et al 2002).

\newpar
The same two 1998 Leonids were also monitored with ELF/VLF radio receivers and there was
no electric ELF/VLF signal above 500 Hz during these two electrophonic events. However,
such signals were detected from other Leonid meteors during the same observational
campaign (Garaj et al 1999). This result is basically confirmed by Shawn E. Korgan from
the NASA INSPIRE Team I-01 (GEFS2001\_11\_18\_08). He was recording the atmospheric VLF
activity when he heard electrophonic sounds from meteors. The recordings did not show
any VLF activity correlated with the sound events. This is consistent with the detection
of geomagnetic disturbances below 10 Hz detected during the reentry of an artificial
satellite accompanied by electrophonic sounds (Verveer, Bland \& Bevan 2000) and with
the electric field disturbances below several hundreds of Hz correlated with the
activity of 2001 Leonids (Trautner et al 2002).

\newpar In 8 occasions, the observers associated meteor fragmentation with an
electrophonic sound. Six of these are very transient in duration: ``pop'', ``boom'', and
``crack''. This suggests a sudden

\par\eject\noindent

\table{{\it Distribution of the electrophonic meteor magnitudes. The rate of occurrence
derived by Kaznev (1994) is also given. It has been argued by other authors that the low
brightness meteors can not produce sustained electrophonic sounds, thus we also show the
sound descriptions.}} {
   \vrule# &
   \hfill\quad\strut#\hfill\quad &
   \vrule# &
   \hfill\quad\strut#\hfill\quad &
   \vrule# &
   \hfill\quad\strut#\hfill\quad &
   \vrule# &
   \hfill\quad\strut#\hfill\quad &
   \vrule# &
   \hfill\quad\strut#\hfill\quad &
   \vrule#\cr
 \noalign{\hrule}
 height4pt&\omit&width0pt&\omit&&\omit &&\omit&&\omit&\cr
          & \multispan3 magnitude && rate by && magnitude && sound descriptions &\cr
 height2pt&\multispan3\leaders\hrule\hfill&&\omit &&\omit&&\omit&\cr
 height2pt&\omit&&\omit&&\omit &&\omit&&\omit&\cr
          & range && rate&& Kaznev && descriptions && \omit &\cr
 height4pt&\omit&&\omit&&\omit &&\omit&&\omit&\cr
 \noalign{\hrule}
 height4pt&\omit&&\omit&&\omit &&\omit&&\omit&\cr
          & -1 to -5 && 36.8\% && 11.3\% && -2 or more, && crackling, sizzling like bacon frying, &\cr
          &\omit&&\omit&&\omit&& not so bright to 2, && sizzling ``sss'', soft hissing, &\cr
          &\omit&&\omit&&\omit&& max -2 to -3,  && hissing followed by a crack, fffffffffp,&\cr
          &\omit&&\omit&&\omit&& -1 with -3 end flare, && short burst of static, short sharp crack, &\cr
          &\omit&&\omit&&\omit&& -1 in twilight,  && broken filament  shaking in blub, &\cr
          &\omit&&\omit&&\omit&& bright, clearly && ``thwuck'', pop, crackling, swoosh, woosh, &\cr
          &\omit&&\omit&&\omit&& visible, bright at && high pitched whistle, fizzing with &\cr
          &\omit&&\omit&&\omit&& Sirius, twice && crackling, loud high-pitched hissing,  &\cr
          &\omit&&\omit&&\omit&& Sirius  && faint hissing, crackled/hissed, pop &\cr
 height4pt&\omit&&\omit&&\omit &&\omit&&\omit&\cr
 \noalign{\hrule}
 height4pt&\omit&&\omit&&\omit &&\omit&&\omit&\cr
          & -5 to -10 && 41.3\% && 19.7\% && very bright, && ``phtt'' like electric arc, lit match,&\cr
          &\omit&&\omit&&\omit&& -5 or so,             && steam escaping from cooker, &\cr
          &\omit&&\omit&&\omit&& -6.5$\pm$0.5$^*$,     && crackling fireworks, sizzling/crackling, &\cr
          &\omit&&\omit&&\omit&& -6 to -7,             && ``sSHheewwuu'', fizzing/hissing, swishing, &\cr
          &\omit&&\omit&&\omit&& -5$\pm$1$^*$,         && hissing, crackling, whuss, pop,  &\cr
          &\omit&&\omit&&\omit&& firework/flare,       && ``sjhhhhh..'', hiss, sizzling ending with pop, &\cr
          &\omit&&\omit&&\omit&& fireball,             && single engine "Cesna" airplane,  &\cr
          &\omit&&\omit&&\omit&& seen in evening,      && like static/crackling,  &\cr
          &\omit&&\omit&&\omit&& -8 to -10,            && ``sss'' with a slight ``zz'' &\cr
          &\omit&&\omit&&\omit&& brighter than Venus   && \omit &\cr
 height4pt&\omit&&\omit&&\omit &&\omit&&\omit&\cr
 \noalign{\hrule}
 height4pt&\omit&&\omit&&\omit &&\omit&&\omit&\cr
          & brigther && 21.9\% && 69.0\% && brighter than && whistling with buzzing, whisper, sizzling, &\cr
          &than -10 &&\omit&&\omit&& the full moon, && rustling like a rocket, wood on fire,  &\cr
          &\omit&&\omit&&\omit&& like full moon,   && white noise, thrumming, lit match,  &\cr
          &\omit&&\omit&&\omit&& bright as moon,   && ``sss'' followed by pop, ''voom'', pop,  &\cr
          &\omit&&\omit&&\omit&& extremely bright,   && whoosh like rustling, hissing/fizzing &\cr
          &\omit&&\omit&&\omit&& lit up the whole && \omit &\cr
          &\omit&&\omit&&\omit&& sky, lit up the && \omit &\cr
          &\omit&&\omit&&\omit&& ground, brightest  && \omit &\cr
          &\omit&&\omit&&\omit&& ever seen, -12$\pm$1$^*$,&& \omit &\cr
          &\omit&&\omit&&\omit&& -15 to -20,  && \omit &\cr
          &\omit&&\omit&&\omit&& -9 to -13  && \omit &\cr
 height4pt&\omit&&\omit&&\omit &&\omit&&\omit&\cr
 \noalign{\hrule}
 height4pt width0pt&\omit&width0pt&\omit&width0pt&\omit&width0pt&\omit&width0pt&\omit&width0pt\cr
 \noalign{$^{*}$Absolute magnitude}
}

\newpar \noindent
release of large amounts of electric charge. Considering the mobility of electrons and
ions, this burst of charge has to be either in excess of electrons or highly anisotropic
(or both) in order to create a net long-range electric field. It remains a mystery,
however, why this process does not happen, or at least not with the same energy scale,
during any other similar meteor fragmentation in nature.

\newpar
Another interesting unusual phenomenon related to an electrophonic fireball is reported
in GEFS1977\_09\_00\_01: a warm ``puff of wind ... towards the end of the duration of
the sound''. Similar tactile phenomena like ``oscillations and shaking of the air''
(Kaznev 1994) or `` oppression of air'' (Romig \& Lamar 1963) have been reported since
the beginning of the history of electrophonic phenomenon. In 1719, Sir Edmund Halley
dismissed ``hearing [meteor's] hiss'' and ``the warmth of its beams'' as ``the effect of
fancy'' (Halley 1719).

\newpar
Appearance of smell simultaneously with a bright meteor has a similar history. There is
one (GEFS1969\_06\_00\_01) GEFS report mentioning a smell of sulphur, one of ozone
(GEFS0000\_11 \_00\_02), and one of ``lightning'' (probably also ozone)
(GEFS1998\_08\_12\_01). Such phenomena

\par\eject\noindent
have been documented in the electrophonic catalogs (Kaznev 1994). The smell of sulphur
and onion was reported during the 1833 Leonids (Olmsted 1833). More recently, a ``foul
metallic, chemical or sulphurous odor'' was reported to accompany the flight of the
Tagish Lake meteorite in 2000 (Brown, ReVelle, \& Hildebrand 2001). These phenomoena are
even more rare than electrophonic sounds. The tactile sensations could be explained by
vibrations of human hair in oscillatory electric fields (Carstensen 1986), while the
smell comes from the ozone production (and some other chemicals) by corona discharge
(Romig \& Lamar 1963, Aubrecht, Stanek \& Koller 2001). Nevertheless, these explanations
remain a speculation since a comprehensive study of those phenomena has never been
performed in the meteor astronomy.

\newpar
{\bf 2.4. Details about the meteor}

\newpar
Thirty eight reported meteors (events) are identified as {\it sporadic} (48\%), 34 as {\it
Leonids} (43\%), 7 as {\it Perseids} (9\%), and one as possible Delta-Aquarid. One of the
Leonids is probably misidentified (GEFS1998\_11\_16\_01) because the radiant was below
the horizon at the time of the event. The range of electrophonic meteor magnitudes
shown in Table~2 is of a special interest for theoretical work since it carries
information about the energetics of electrophonic events. The range of magnitudes is
divided into three groups: between -1$^m$ and -5$^m$, between -5$^m$ and -10$^m$, and
-10$^m$ or brighter. Sometimes it is not easy to make a magnitude estimate; thus, we
provide their descriptions to show our method. The distribution is compared with the
statistical results of Kaznev (1994) who had a sample of 71 electrophonic meteors with
known magnitudes.

\newpar
Our results are clearly different from Kaznev's distribution. Almost 80\% of our meteors
are not brighter than -10, compared to about 30\% by Kaznev. This suggests that our
survey is far less biased toward extremely bright meteors, in contrast to all other
existing electrophonic catalogues. This is understandable because most of their
electrophonic meteors were extracted from catalogues (or reports in the literature) of
very bright fireballs. From the theoretical point of view, it is very interesting that
the lower brightness limit for electrophonic meteors can be as low as approximately
-2$^m$. One can argue that these meteors can have much brighter absolute magnitude, but
their height above horizon clearly shows that this is not the case (one of them is also
photographed, see next section). Keay (1992) (see also Keay 1994) argues, in the context
of his theory, that electrophones from the -7$^m$ or fainter meteors should be very
transient in nature, lasting for a tenth of a second or so. Again, the reports shown in
Table 2 demonstrate that this is not the case for many of such sounds.

\newpar
The velocity of meteors is described as {\it very slow} in 5 reports (6\%), as {\it
slow} in 38 reports (42\%), as {\it fast} in 40 reports (45\%), 5 as {\it very fast}
(6\%), and one meteor as {\it stationary} (1\%). Meteor fragmentation is reported for
32 events (38\%) and it did not occur in 53 events (62\%). The distribution of meteor
height above horizon, its azimuth, and angle between its path and horizon is shown in
Table 3. For comparison, distribution from Kaznev (1994) is also shown. Our statistical
sample is big enough to notice some interesting statistical averages.

\newpar
The distribution of height above horizon of the electrophonic meteors from Kaznev peaks
with about 45\% in the 30-60$^o$ region. Our survey shows only 30\% of meteors in this
region. However, 45\% of our meteors are above 60$^o$, while Kaznev reports only 25\%.
Even though people tend to overestimate this angle, this mismatch is significant because
such overestimates appear in both surveys and they are statistically averaged. This
suggests that something else is responsible for shifting our distribution closer toward
the zenith.

\newpar
We propose two explanations. The first explanation is that a larger number of smaller
meteors in our sample. Indeed, there is a slight increase in the angle for the -5$^m$ to
-10$^m$ meteors compared to the -1$^m$ to -5$^m$ meteors, but the statistical
uncertainty is too large for any conclusive differentiation. The second explanation is
that smaller meteors can not produce a very strong EM signal. This would imply that they
have to be closer to the observer, that is closer to the zenith. Since all of our
meteors are bright enough to be visible from a large distance, this explanation seems
plausible.

\par\eject\noindent
However, the azimuthal angle shows a very random distribution of observers around
meteors. One quarter of meteors appear in each of four 90$^o$ intervals, which is also
noticeable in Kaznev's distribution. The angle between the meteor path and horizon also
shows similarity to Kaznev's results, with approximately 50\% of meteors with available
data in the 0-30$^o$ region, 25\% in the 30-60$^o$ region, and 25\% over 60$^o$.

\newpar\noindent
\table{{\it Statistical analysis of the meteor path in the sky. An event represents one
observing site. If an angle is not clear cut between two statistical regions (e.g 30$^o$
for the hight above horizon) then the event is counted as 0.5 in both adjacent regions,
or 0.33 when spanning over three regions. The results are compared to the values by
Kaznev (1994).}} {
   \vrule# &
   \hfill\quad\strut#\hfill\quad &
   \vrule# &
   \hfill\quad\strut#\hfill\quad &
   \vrule# &
   \hfill\quad\strut#\hfill\quad &
   \vrule# &
   \hfill\quad\strut#\hfill\quad &
   \vrule# &
   \hfill\quad\strut#\hfill\quad &
   \vrule#\cr
 height2pt width0pt&\omit&width0pt&\multispan7\leaders\hrule\hfill&width0pt\cr
 height2pt width0pt&\omit&&\omit &&\omit&&\omit&&\omit&\cr
   width0pt & \omit && Angle && Events && Rate  && Rate by & \cr
   width0pt & \omit && (deg) && \omit  && \omit && Kaznev  & \cr
 height2pt width0pt&\omit&&\omit &&\omit&&\omit&&\omit&\cr
 \noalign{\hrule}
 height2pt&\omit&&\omit &&\omit&&\omit&&\omit&\cr
   & Height above horizon && 0-30 && 18.7 && 23.9\%    && 31.6\% & \cr
 height2pt&\omit&&\omit &&\omit&&\omit&&\omit&\cr
   & \omit                && 30-60 && 23.8 && 30.6\%   && 43.9\% & \cr
 height2pt&\omit&&\omit &&\omit&&\omit&&\omit&\cr
   & \omit                && 60-90 && 35.5 && 45.5\%   && 24.5\% & \cr
 height2pt&\omit&&\omit &&\omit&&\omit&&\omit&\cr
 \noalign{\hrule}
 height2pt width0pt&\omit&width0pt&\omit &width0pt&\omit&width0pt&\omit&width0pt&\omit&width0pt\cr
 \noalign{\hrule}
 height2pt&\omit&&\omit &&\omit&&\omit&&\omit&\cr
   & Azimuth              && 315-45(N) && 14.3 && 25.6\%   && 22.4\% & \cr
 height2pt&\omit&&\omit &&\omit&&\omit&&\omit&\cr
   & \omit                && 45-135(E) && 15.3 && 27.4\%   && 28.5\% & \cr
 height2pt&\omit&&\omit &&\omit&&\omit&&\omit&\cr
   & \omit                && 135-225(S) && 15.8 && 28.3\%   && 22.3\% & \cr
 height2pt&\omit&&\omit &&\omit&&\omit&&\omit&\cr
   & \omit                && 225-315(W) && 10.5 && 18.8\%   && 26.8\% & \cr
 height2pt&\omit&&\omit &&\omit&&\omit&&\omit&\cr
 \noalign{\hrule}
 height2pt width0pt&\omit&width0pt&\omit &width0pt&\omit&width0pt&\omit&width0pt&\omit&width0pt\cr
 \noalign{\hrule}
 height2pt&\omit&&\omit &&\omit&&\omit&&\omit&\cr
   & Angle between        && 0-30 && 25.7 && 52.4\%   && 44.5\% & \cr
 height2pt&\omit&&\omit &&\omit&&\omit&&\omit&\cr
   & the meteor path      && 30-60 && 11.7 && 23.8\%   && 32.5\% & \cr
 height2pt&\omit&&\omit &&\omit&&\omit&&\omit&\cr
   & and horizon          && 60-90 && 11.7 && 23.8\%   && 23.0\% & \cr
 height2pt&\omit&&\omit &&\omit&&\omit&&\omit&\cr
 \noalign{\hrule}
}

\section{Reports of special interest}

A couple of reports attract special attention either because they have been extensively
documented by observers or they deal with an interesting type of meteor. We present
details about the electrophonic sounds from Leonids, a photo of one low brightness
electrophonic meteor, and meteors with estimated trajectory. More about all these events
can be found at the GEFS homepage.

\newpar
{\bf 3.1. Electrophonic sounds from Leonids}

\newpar
The Leonids, and meteors with similar properties like Perseids, are the biggest
theoretical challenge in explaining the electrophonic phenomenon. Not only are there low
magnitude electrophonic Leonids which disintegrate at altitudes above 80 km, but there
are also sustained sounds from the Leonids. A sustained sound should last for a large
fraction of a second in order to be perceived as such by the observer. After taking into
account their high velocity, we see that the electrophonic signal can start at
exceptionally high altitudes of $\sim$100 km. These altitudes have been also obtained by
the instrumental recordings of electrophonic sounds from the 1998 Leonids (Zgrabli\'{c}
et al 2002).

\newpar
Altogether there are 34 reports of sounds from the Leonids. One report is about the 1964
Leonids, two about 1966, one about 1989, 10 about 1998, one about 2000, 17 about 2001,
and two are without a specific year. The sound duration is usually overestimated by the
witnesses, thus durations of $\sim$3 seconds are not surprising. The sound description
spans from high-frequency sounds like ``hissing'', ``sizzling'', ``crackling'',
``fizz'', ``swoosh'', or ``white noise'', to low-frequency sounds like ``(deep) pop'',
``boom/popping'', or ``clap''. The magnitudes range from as low as -2$^m$ to ``bright
enough to light up the ground'' or ``the whole sky''. One case of a -2$^m$ meteor is
also described in Drummond, Gardner \& Kelley (2000) (GEFS1998\_11\_17\_01).

\par\eject\noindent
Details about GEFS1998\_11\_17\_04 and GEFS1998\_11\_17\_05 are available in Zgrablic et
al (2002). As already mentioned above, the VLF radio signal did not accompany these two
electrophonic meteors, as it did not the meteors in GEFS2001\_11\_18\_08.

\newpar
The reports GEFS1966\_11\_17\_01 and GEFS1966\_11\_17\_02 represent the first documented
report of electrophonic sounds known to us from the famous Leonid meteor storm of 1966.
The first observation took place in Texas, USA, from about 5:30 a.m. until 7:00 a.m.
local time, when the radiant was 70-80$^o$ above the horizon. According to the witness
Willis Jarrel Jr., ``the sounds came intermittently from the beginning of the
observation until the end''. It was not possible to connect particular meteors with the
sounds, except in one case of an extremely bright fireball. This demonstrates again the
existence of low magnitude electrophonic Leonids. The sounds were lacking
directionality.

\newpar
They are described as ``a velvet silky rustling sound like a lady walking in a pleated
dress where the fabric rubs against itself'' and ``some sounded like a short distorted
hiss, with a pronounced sibilant tone''. These sounds had shorter duration than the one
connected to the bright fireball and were described as similar to the other sounds but
``lower in pitch and much more edgy and crackling''. The witness notes that he has
better than average hearing, which explains his experience of a large number of
electrophonic sounds. The witness has also provided photos of the observation site
(second-story open-air deck on a house). The photos and additional details about the
event are available on the GEFS homepage.

\newpar
He also notes that he woke up and went to a window for no reason, probably because of a
``stimulus of some sort''. Even though the existence of a ``stimulus'' sounds
unrealistic, this is not a unique report of this sort (Kaznev 1994) and can not be
ignored. Possible physical explanation could be that the witness was exposed to frequent
bursts of strong electric fields, as implied by the large number of electrophonic
sounds. According to laboratory experiments, animals, especially, and humans can be
sensitive to the short pulses of electric fields (Buskirk, Frohlich \& Latham 1981).
Thus the reality of such ``stimulus'' remains an open question for future research.

\newpar
The second observation of the 1966 Leonids was from Kansas, USA, from about 1:00 a.m. to
4:00 a.m. local time. The witness recalls hearing approximately 20 ``noisy'' meteors
that night. They sounded like ``an electric flutter or sizzle'' with one half of a
second duration. The magnitudes are described just as ``all magnitudes''.

\newpar
{\bf 3.2. Photo of a low magnitude electrophonic meteor}

\newpar
Electrophonic meteors with magnitudes as low as $-2^m$ make a significant fraction of
the electrophonic sound reports (see Table 2). Since they represent a challenge to the
theoretical modeling of the phenomenon, here we present a photo of one of them.

\newpar
The report GEFS1972\_04\_23\_01 belongs to Eisse Pieter Bus from The Netherlands who was
performing visual and photographic meteor observations on the night of April 22/23,
1972, at the Observatory of the University of Groningen at Roden. At 01:12:47UT, a
$-2^m$ meteor passed through the constellation of Corona Borealis. The meteor was
photographed by the camera during a one minute exposure time (see Figure 1). Since
Corona Borealis was $\sim$65$^o$ above the horizon at that time, the absolute meteor
magnitude was close to the estimated apparent $-2^m$.

\newpar
The witness recalls hearing a cracking sound during the whole flight of the meteor in
duration of $\sim$5 seconds. The sound did not have direction. ``It was in the middle of
[his] head like a sound in a stereo headphone''. The meteor did not have a light
maximum, but it showed ``a wake that moved slowly from the left to the right (about
20$^o$ to the left and right). Close behind this wake, but not connected, a persistent
trail was visible with a lifetime of about 1 second''. The witness emphasized that he
has ``seen hundreds of bright ... and very bright meteors but [he has] never heard a
meteor with a sound nor [he has] seen a meteor with a wake again''.

\par\eject\noindent

\fullfigure{15truecm}{10truecm}{ {\it A $-2^m$ electrophonic meteor on 23 April 1972, at
01:12:47UT photographed from the Observatory of the University of Groningen at Roden,
The Netherlands. The meteor's direction is from South to North, the exposure time was
01:12:17UT - 01:13:17UT with Exa 1a 2.8/50 mm camera and Kodak-Tri-X film with 27$^o$
DIN. Courtesy of Eisse Pieter Bus.} } {\includegraphics{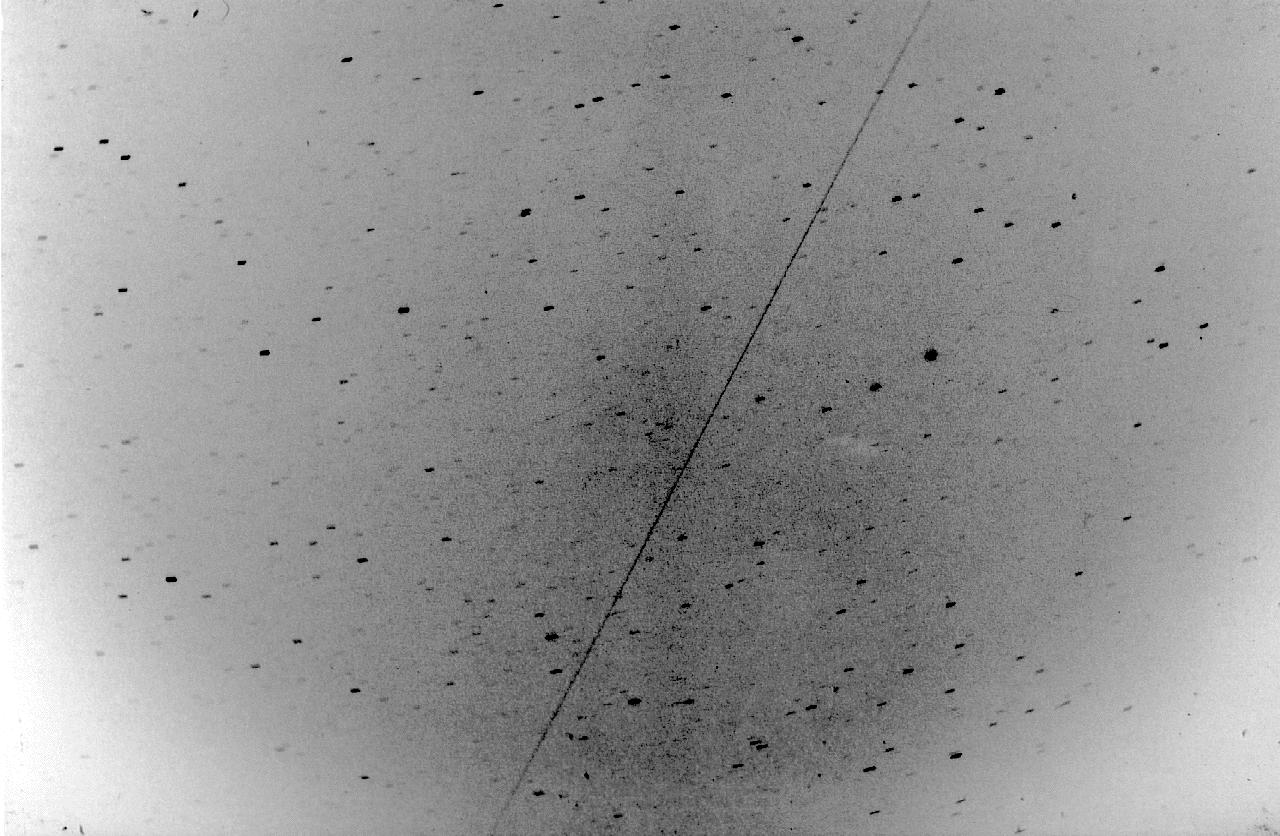}}

\newpar
{\bf 3.3. Fireball over north England, 9 January 2000}

\newpar
The witness reports of this event were collected by Alastair McBeath, who analyzed them
and posted the results to the IMO-News e-mail list. All the information presented here
are part of these results and published in McBeath (2000). The electrophonic event is
cataloged as GEFS2000\_01\_09\_01.

\newpar
The fireball occured on 9 January 2000, over north England, UK, at around 01:56 UT.
The estimated visible trajectory starts above Appleby in Cumbria
(\hbox{54$^{\circ}$35'N}, \hbox{02$^{\circ}$30'W}) and ends \hbox{$\sim$10 km} offshore
due east of Seaton Sluice, Northumberland (\hbox{55$^{\circ}$05'N},
\hbox{01$^{\circ}$15'W}). The entry angle was 33$\pm$3$^{\circ}$ from the horizonal,
which gives the atmospheric path length of approximately $\sim$110 km, with the mean
atmospheric velocity of 22$\pm$3 km/s. The estimated  brightness was between -15$^m$ and
-20$^m$.

\newpar
There was several reports of acoustic signals, one of which is recognized as an
electrophonic sound. A whoosh sound, ``like a rustling'', was reported from an observer
located on top of a hill called Eston Nab (\hbox{54$^{\circ}$33'30''N},
\hbox{01$^{\circ}$07'W}), at the closest distance of approximately $\sim$60 km
south-east from the ground track. It is interesting that the noise was associated with
the breaking up during the flight when ``three large lumps, glowing like red-hot brick''
separated off the main body, two of which were significantly smaller than the third.

\newpar
The Earth's magnetic field in the vicinity of the meteor is useful information for a
future theoretical work. The magnetic field components (National Geophysical Data
Center, The World Data Center for Solid Earth Geophysics, Boulder,
http://www.ngdc.noaa.gov/seg/wdca/) on that day at location 55$^{\circ}$N 02$^{\circ}$W
and 50 km altitude are Z=44,994 nT (vertical, direction down), H=17,181 nT (horizontal),
with magnetic declination of \hbox{4$^{\circ}$46'W} (model IGRF2000). Variations from
these values along the fireball path are $\sim$1\% or less.

\par\eject\noindent
\fullfigure{12truecm}{13truecm}{ {\it Electrophonic fireball over Denmark on 20 December
1999. The ground track is a rough estimate. The electrophonic sound events are marked by
points and numbers. See text for their description and more details about the fireball.
Courtesy of Holger Pedersen.} } {\includegraphics{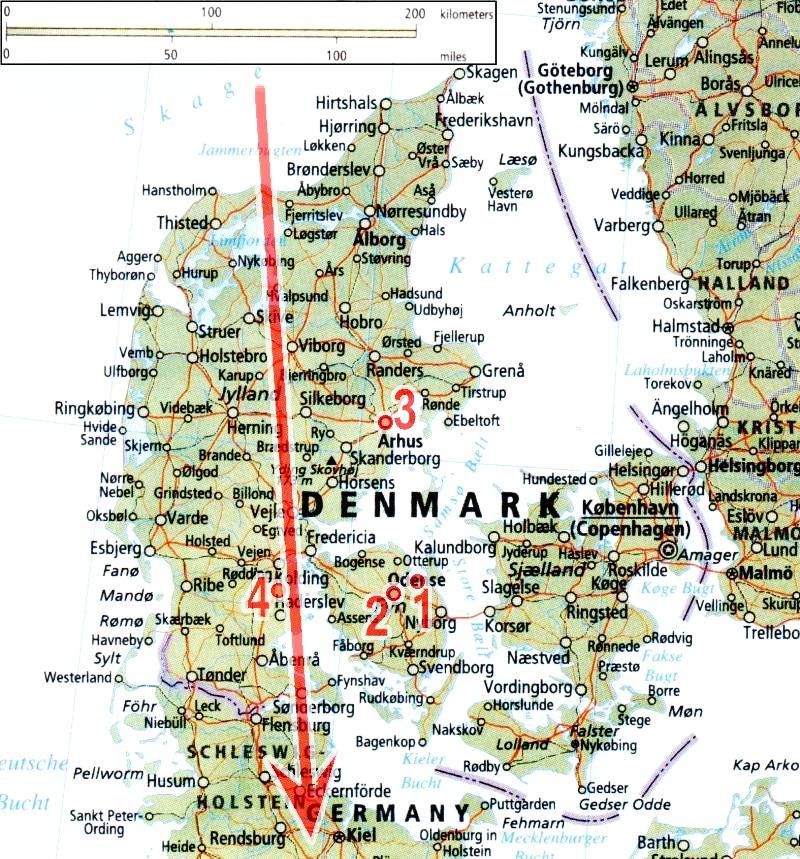}}

\newpar
{\bf 3.4. Fireball over Denmark, 20 December 1999}

\newpar
The witness reports of this event were collected by the Tycho Brahe Planetarium,
Copenhagen and provided to Holger Pedersen by its director Bjoern Franck Joergensen.
The information presented here was obtained from a statement by the Planetarium and
from the IMO-News e-mail list, where several e-mails related to the event were posted.
Additional details about the electrophonic sound report were provided to the GEFS by
Holger Pedersen and cataloged as GEFS1999\_12\_20\_01.

\newpar
The fireball occured on 20 December 1999, over Denmark, at around \hbox{19:15 UT}. The
exact trajectory is not determined. According to Lars Bakmann (Meteor Section
Astronomical Society, Denmark) the meteor was passing zenith above S{\o}nderborg
(\hbox{54$^{\circ}$54'N}, \hbox{09$^{\circ}$47'E}) with the azimuth of
170$\pm$20$^{\circ}$ (direction from the north to south). The azimuth favors larger
angles, since the fireball was visible from G{\"{o}}teborg (Sweden) and the Oslo area
(Norway). The trajectory was very shallow, often described as "almost parallel to the
horizon". The altitude is uncertain. If the visible part of the flight started at
an altitude of $\sim$110 km over the sea between Denmark and Norway and terminated at
an altitude of $\sim$40 km above the region of the town Kiel in Germany, the ground track would
be $\sim$400 km, and the angle of flight would be $\sim$10$^{\circ}$ with the
horizontal. The mean atmospheric velocity was $\sim$10 km/s. All these numbers are
rough estimates, including the meteor's magnitude of -5$\pm$1$^m$.

\par\eject\noindent
Five witnesses at four different locations reported acoustic signals recognized as
electrophonic sounds:

\noindent {\bf (1)} observer from Munkebo (\hbox{55$^{\circ}$27'N}, \hbox{10$^{\circ}$34'E})
heard ``a subdued, hissing sound ... (like) a boat which gently
slides through water'';

\noindent {\bf (2)} observers from Odense (\hbox{55$^{\circ}$24'N},
\hbox{10$^{\circ}$23'E}) heard a hissing sound when ``a couple of small peaces
detached'';

\noindent {\bf (3)} observer from {\AA}rhus (\hbox{56$^{\circ}$09'N},
\hbox{10$^{\circ}$13'E}) heard a faint hiss;

\noindent {\bf (4)} and observer from Christiansfeld
(\hbox{55$^{\circ}$21'N}, \hbox{09$^{\circ}$29'E}) also heard a hiss.

\newpar
The meteor ground track and location of electrophonic events is shown in Figure 2. The
magnetic field components at \hbox{56$^{\circ}$N} \hbox{10$^{\circ}$E} and 50 km
altitude are Z=45,700 nT (vertical, direction down), H=16,594 nT (horizontal), with
magnetic declination of \hbox{0$^{\circ}$01'E} (model IGRF95).

\newpar
{\bf 3.5. Fireball over Croatia, 3 November 1997}

\newpar
The witness reports of this event were collected by Korado Korlevi\'{c}, Vi\v{s}njan
Observatory, Croatia, and the information presented here is the result of his
analysis. When interviewing the witnesses (usually by phone), he recognized
electrophonic sound events on several occasions and made a note about their location and
the name of the observer but no other details. This fireball is cataloged as
GEFS1997\_11\_03\_01.

\newpar
\fullfigure{16truecm}{12.8truecm}{
{\it Electrophonic fireball over Croatia on 3 November 1997.
The electrophonic sound events are marked by points and numbers. Their
locations are:
{\bf (1)} Pula (\hbox{44$^{\circ}$52'N}, 13$^{\circ}$51'E),
{\bf (2)} Pore\v{c} (\hbox{45$^{\circ}$13'N}, 13$^{\circ}$36'E),
{\bf (3)} Zadar (\hbox{44$^{\circ}$07'N}, 15$^{\circ}$15'E),
{\bf (4)} Duga Resa (\hbox{45$^{\circ}$27'N}, 15$^{\circ}$30'E),
{\bf (5)} Josipdol (\hbox{45$^{\circ}$12'N}, 15$^{\circ}$17'E),
{\bf (6)} Dabar (\hbox{44$^{\circ}$57'N}, 15$^{\circ}$19'E).
See text for more details about the fireball. Courtesy of Korado Korlevi\'{c}.}}
{\includegraphics{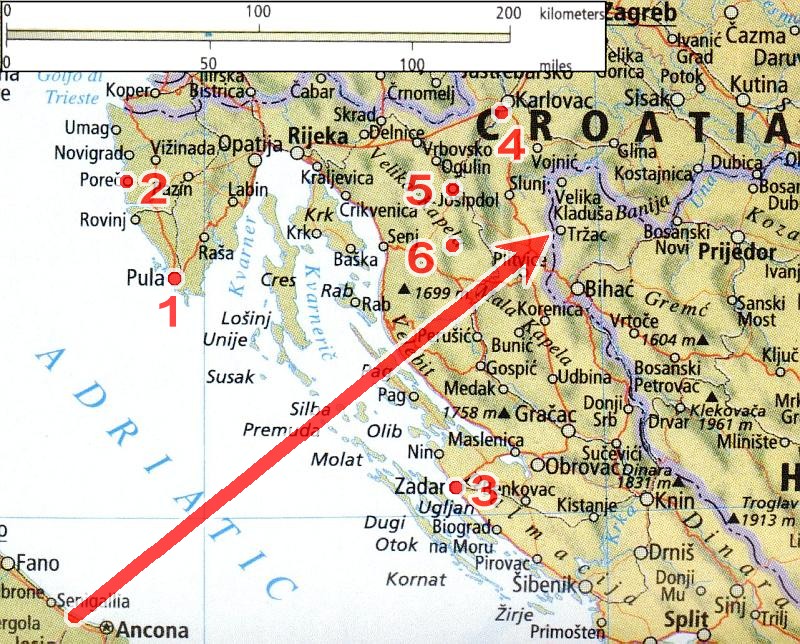}}

\par\eject\noindent
The fireball occurred on 3 November 1997, over the Adriatic Sea and Croatia at 16:08:20
UT. The estimated visible trajectory starts over the Italian Adriatic coast
(\hbox{43$^{\circ}$43'N}, \hbox{13$^{\circ}$13'E}), close to Ancona, and ends over the
border between Croatia (CRO) and Bosnia and Herzegovina (BH) (\hbox{44$^{\circ}$59'N},
\hbox{15$^{\circ}$45'E}) at an altitude of 30-40 km. The angle between the trajectory and
horizontal is $\sim$15$^{\circ}$ with $\sim$250 km of ground track. The meteor displayed
multiple fragmentation over Velebit mountain. Fragments burned out quickly except for
one of them which continued the flight parallel to the main body. The final
fragmentation happened between Dre\v{z}nik Grad (CRO) and Tr\v{z}ica (BH) with rapid
deceleration (duration of the final flight was 3-4 seconds). The mean atmospheric
velocity, excluding the final deceleration, was 20-25 km/s. Witnesses described the
meteor as brighter than a full Moon.

\newpar
The electrophonic sounds were reported from six different locations. The sounds are
described as rustling or ``like a rocket'', but there are no details about particular
events. The meteor ground path and locations of electrophonic events are shown in Figure
3. The magnetic field components at \hbox{44$^{\circ}$N} \hbox{14$^{\circ}$E} and 50 km
altitudes are Z=39,522 nT (vertical, direction down) and H=22,618 nT (horizontal), with
magnetic declination of 1$^{\circ}$18'E (model IGRF95).

\section{Conclusion}

The analysis described in this study revealed two important facts: i) electrophonic
sounds can appear even for meteors of a visual magnitude lower than previously thought,
and ii) the estimated heights where electrophonic meteors enter the atmosphere can reach
high values (even 100 km) which has implications on the theories of meteor ELF/VLF
generation. From the theoretical standpoint, these new facts demonstrate that very
little has changed since the early work in this field in the 1960's (Bronshten 1991).

\newpar
Moreover, it has become widely accepted that the electrophonic sounds are created
exclusively by vibration of ordinary objects exposed to the ELF/VLF electric fields,
even though there are experiments which show corona discharge with the same value of
electric fields. Thus, the catalogs like GEFS are still very useful and can be used for
testing the existing theories.

\newpar
The most important result, coming from the GEFS witness reports, is the lower limit on the
magnitude of electrophonic meteors. The catalogues of electrophonic sounds studied so
far have been observationally biased toward very bright fireballs, since such meteors
are often individually studied and attract a lot of attention. The brightness limit
often cited in the literature is about -10$^m$ for sustained sounds and about -7$^m$ for
more transient sounds (Keay 1992, Beech \& Foschini 1999). However, Kaznev's
analysis of electrophonic meteors already pointed toward the existence of sounds from
meteors of magnitude as low as -2$^m$. The GEFS reports show that such low brightness
electrophonic meteors (dimmer than -7$^m$) really exist and represent a large fraction
of the electrophonic sound events; moreover, they can produce sustained sounds instead
of only transient sounds.

\newpar
It is also important to notice that there are Leonids among these low brightness
meteors. They ablate at very high altitudes, and sustained sounds from them indicate that
the electrophonic effects may already start to appear at altitudes of about 100km. These
are also altitudes of the beginning of nighttime ionosphere. This is consistent with the
instrumental recording of the electrophonic sounds from the 1998 Leonids (Zgrablic at al
2002). An increase of height above the horizon of meteors in the GEFS reports, compared to
Kaznev's results, could indicate that the EM effects from low brightness meteors are not
as strong as from very bright fireballs. Presented examples of bright fireballs with the
known trajectory show that the electrophonic sounds can be induced even at distances
over 100km from the fireball's ground track.

\newpar
The results presented here are a big challenge for the theory. Any future work will
require more expreimental/observational results and multidisciplinary research.

\par\eject\noindent
\appsection{Acknowledgments}%
We thank the Center for Computational Sciences at the University of Kentucky and the Swiss
Federal Institute of Technology Lausanne (EPFL) for providing technical support for
GEFS. We are grateful to all the people who sent their witness reports to GEFS. We
gratefully acknowledge the extensive data and materials provided by Eisse Pieter Bus,
Alastair McBeath, Holger Pedersen, Korado Korlevi\'{c}, and Willis Jarrel Jr. We also
thank Cora Allard and Helen Klarich for their assistance in preparing this paper.
\appsection{References}%
 \bookref{L.~Aubrecht, Z.~Stanek, J.~Koller}{Corona discharge on coniferous trees - spruce
 and pine} {Europhys. Lett. 53, 2001, pp.~304-309}
 \bookref{M.~Beech, L.~Foschini}{A space
 charge model for electrophonic bursters} {Astron.~Astrophys. 345, 1999, pp.~L27-L31}
 \bookref{V.A.~Bronshten}{Electrical and electromagnetic phenomena associated with the meteor
 flight}{Sol.~Syst.~Res. 25, 1991, pp.~93-104}
 \bookref{P.~Brown, D.O.~ReVelle, A.R.~Hildebrand}{The Tagish Lake Meteorite Fall :
 Interpretation of Fireball Physical Characteristics}{ESA SP-495,
 Proceedings of the Meteoroids 2001 conference, Ed. B. Warmbein, 2001, pp. 497-507}
 \bookref{R.E.~Buskirk, C.~Frohlich, G.V.~Latham}{Unusual animal behavior before
 earthquakes: a review of possible sensory mechanisms}{Rev. Geophy. Space Phy. 19, 1981,
 pp.~247-270}
 \bookref{E.L.~Carstensen}{Biological effects of transmission line fields}
 {Proceedings of a Special Symposium on Critical Emerging Issues in Biomedical Engineering,
 IEEE, Ed. C. J. Robinson and G. V. Kondraske, New York, 1986}
 \bookref{P.L.~Dravert}{Electrophonic bolides of Western Siberia}{Byull.~Tsentr.~Komis.~po
 Meteoram, Kometam i Asteroidam Astrosoveta, Akad. Nauk SSSR, 18, 1940, pp.~1-2}
 \bookref{J.D.~Drummond, C.S.~Gardner, M.C.~Kelley}{Catching a falling star}{Sky\&Telescope
  99, June 2000, pp.~46-49}
 \bookref{S.~Garaj et al}{Observational detection of meteor-produced VLF electromagnetic
 radiation}{FIZIKA 8, 1999, pp.~91-98}
 \bookref{E.~Halley}{An account of the extraordinary meteor seen all over England, on the
 19th of march 1718/9. With a demonstration of the uncommon height thereof}
 {Phil.~Trans.~Roy. Soc.~London 30, 1719, pp.~978-990}
 \bookref{V.Yu.~Kaznev}{Observational characteristics of electrophonic bolides: statistical
 analysis}{Sol.~Syst.~Res. 28, 1994, pp.~49-60}
 \bookref{C.S.L.~Keay}{Anomalous sounds from the entry of meteor fireballs}{Science 210, 1980,
 pp.~11-15}
 \bookref{C.S.L.~Keay}{Electrophonic sounds from large meteor fireballs}{Meteoritics 27,
 1992, pp.~144-148}
 \bookref{C.S.L.~Keay}{Electrophonic Sounds Catalog}{WGN Observational Report Series
 6, 1993a, pp.~151-172}
 \bookref{C.S.L.~Keay}{Electrophonic Meteor Fireballs Require Further Study}{Meteoroids and their
 parent bodies, Astronomical Inst., Slovak Acad. Sci., Bratislava, 1993b, Ed. J. Stohl and
 I.P. Williams, pp.315-318}
 \bookref{C.S.L.~Keay}{Audible fireballs and geophysical electrophonics}
 {Proc. Astro. Soc. Australia 11, 1994, pp.~12-15}
 \bookref{C.S.L.~Keay, P.M.~Ostwald}{A laboratory test of the production of electrophonic
 sounds}{J.~Acoust.~Soc.~Am. 89, 1991, pp.~1823-1824}
 \bookref{A.~McBeath}{SPA meteor section results: January-February 2000}{WGN, Journal of the
 International Meteor Organization 28, 2000, pp.~232-236}
 \bookref{D.~Olmsted}{Observations on the meteors of Nov. 13th 1833}
 {The~Acoust.~Soc.~Am. 89, 1991, pp.~1823-1824}
 \bookref{M.F.~Romig, D.L.~Lamar}{Anomalous sounds and electromagnetic effects associated with
 fireball entry}{RAND Memo., RM-3724-ARPA, 1963}%
 \bookref{S.S.~Stevens}{On hearing by electrical stimulation}{J.~Acoust.~Soc.~Am. 8,
 1937, pp.~191-195}%
 \bookref{R. Trautner et al}{ULF-VLF Electric field measurements during the 2001 Leonid
 storm}{Proc. ACM Berlin, 29 Jul - 02 Aug 2002, ESA-SP-500, submitted}
 \bookref{A. Verveer, P.A. Bland, A.W.R. Bevan}{Electrophonic Sounds from the Reentry of the
 Molniya 1-67 Satellite over Australia: Confirmation of the Electromagnetic Link}{Meteorit. Planet.
 Sci. 35 Suppl., 2000, pp.~A163.}
 \bookref{D.~Vinkovi\'{c} et al}{Global Electrophonic Fireball
 Survey}{ WGN, Journal of the International Meteor Organization 28, 2000, pp.~48-53}
 \bookref{D.Y. Wang, T.F. Tuan, S.M. Silverman}{A note on anomalous sounds from meteor
 fireballs and aurorae}{J.~Roy. Astron. Soc.~Can. 78, 1984, pp.~145-150}
 \bookref{G.~Zgrabli\'{c} et al}{Instrumental recording of electrophonic sounds from
 Leonid fireballs}{J.~Geophys.~Res.~-~Space Phys. 107(A7), 10.1029/2001JA000310, 2002}

\appsection{Authors' address}%
 {\it Dejan Vinkovi\'{c}\/}, Department of Physics and Astronomy, University of Kentucky,
Lexington, KY 40506-0055, USA, e-mail: {\tt dejan@pa.uky.edu}.\smallnewpar
 {\it Slaven Garaj\/}, I.G.A.- D\'{e}partement de Physique, Ecole Polytechnique F\'{e}d\'{e}rale
 de Lausanne, CH-1015 Lausanne-EPFL, Switzerland.\smallnewpar
 {\it Pey Lian Lim\/}, Department of Physics and Astronomy, University of Kentucky,
Lexington, KY 40506-0055, USA.\smallnewpar
 {\it Damir Kova\v{c}i\'{c}\/}, Cognitive Neuroscience Sector, International School for Advanced
 Studies, SISSA, via Beirut 2-4, 34014 Trieste, Italy.\smallnewpar
 {\it Goran Zgrabli\'{c}\/}, Institut de Physique de la Matiere Condensee,
Universite de Lausanne, CH-1015 Lausanne, Switzerland.\smallnewpar
 {\it \v{Z}eljko Andrei\'{c}\/}, Department of Materials Science, Thin Films Laboratory,
 Rudjer Bo\v{s}kovi\'{c} Institute, Bijeni\v{c}ka 54, HR-10000 Zagreb, Croatia. \smallnewpar

\par\vfil\eject\end